\documentclass[JHEP,11pt]{article}
\usepackage{verbatim}

\usepackage{dsfont}
\usepackage{sidecap}
\sidecaptionvpos{figure}{t}
\usepackage[utf8]{inputenc}
\usepackage[english]{babel}
\usepackage{empheq}
\usepackage{amssymb,amsmath}
\usepackage{graphicx, xcolor, varwidth}
\usepackage{setspace}
\usepackage[permil]{overpic}
\usepackage{graphicx}
\usepackage{color}
\usepackage{braket}
\usepackage{simplewick}
\usepackage{jheppub}
\usepackage[T1]{fontenc}
\usepackage[autostyle]{csquotes}
\usepackage{ulem}
\usepackage{float}
\usepackage{caption}
\usepackage{subcaption}
\usepackage{mathtools}

\usepackage[bottom]{footmisc}

\usepackage{simpler-wick}

\usepackage{feynmp}
\colorlet{darkblue}{blue!70!black}
\colorlet{darkgreen}{green!70!black}

\numberwithin{equation}{section}

\numberwithin{equation}{section}

\newcommand{\wb}{\bar{w}}

\newcommand{\mA}{\mathcal{A}}

\newcommand{\sgn}{\text{sgn}}

\newcommand{\mF}{\mathcal{F}}

\newcommand{\mJ}{\mathcal{J}}

\newcommand{\mC}{\mathcal{C}}
\newcommand{\disc}{\text{disc}}
\newcommand{\bl}{\boldsymbol{\ell}}
\newcommand{\bi}{\boldsymbol{i}}



		\title{	{Quantum Error Correction in SYK and Bulk Emergence}}

			\author[1]{Venkatesa Chandrasekaran}
			\author[1]{and Adam Levine}

		\affiliation[1]{Institute for Advanced Study, Princeton, NJ 08540, USA}
	
		\emailAdd{venchandrasekaran@ias.edu}
		\emailAdd{arlevine@ias.edu}

		\abstract{We analyze the error correcting properties of the Sachdev-Ye-Kitaev model, with errors that correspond to erasures of subsets of fermions. We study the limit where the number of fermions erased is large but small compared to the total number of fermions. We compute the \emph{price} of the quantum error correcting code, defined as the number of physical qubits needed to reconstruct whether a given operator has been acted upon the thermal state or not. By thinking about reconstruction via quantum teleportation, we argue for a bound that relates the price to the ordinary operator size in systems that display so-called \emph{detailed size winding} \cite{Nezami:2021td}. We then find that in SYK the price roughly saturates this bound. Computing the price requires computing modular flowed correlators with respect to the density matrix associated to a subset of fermions. We offer an interpretation of these correlators as probing a quantum extremal surface in the AdS dual of SYK. In the large $N$ limit, the operator algebras associated to subsets of fermions in SYK satisfy half-sided modular inclusion, which is indicative of an emergent Type III$_1$ von Neumann algebra. We discuss the relationship between the emergent algebra of half-sided modular inclusions and bulk symmetry generators.}

\addtocounter{page}{1}

\begin{document}

\maketitle 

\section{Introduction}\label{sec:intro}

Quantum error correction is intimately linked to the appearance of gravity, geometry and bulk locality from holographic quantum mechanical systems, as was emphasized in the pioneering work of  \cite{Almheiri:2014lwa,Dong:2016eik,Harlow:2016vwg}. On the other hand, much work has been done to understand the emergence of a bulk dual in terms of operator size growth and complexity \cite{Susskind:2019ue,Brown:2018wp,Nezami:2021td,Haehl:2021tx, Qi:2018wg,Schuster:2021uc}. In this work, we attempt to relate these two concepts, using the Sachdev-Ye-Kitaev (SYK) model as an illustrative example \cite{Maldacena:2016vs,kitaevsyk,Sachdev:1993tw}.

We start by asking a simple question, which has been asked before in many contexts \cite{Hayden:2007wr}: when can a given operator acting on a quantum system in thermal equilibrium be reconstructed on some subset of that system? In the context of a quantum mechanical system of spins, such as SYK, one might ask about scrambling dynamics. If we act with a Pauli operator at some time $t=0$ and then evolve the resulting state forward in time, is it possible to reconstruct the action of this Heisenberg-evolved operator on some subset of spins? How big must this subset be?

In the precise language of quantum error correction, we can imagine a small code subspace, say one qubits worth, which encodes whether the thermal state has been disturbed by the action of a Heisenberg evolved operator or not. In other words, we imagine a two-dimensional code subspace spanned by the two states 
\begin{align}\label{eqn:Hcode}
    \mathcal{H}_{code}(T_R) = \text{span} \lbrace \ket{\beta}_{LR},\ \psi(T_R)\ket{\beta}_{LR}\rbrace,
\end{align}
where $\psi(T_R) = e^{iH_R T_R} \psi e^{-iH_R T_R}$ and $\ket{\beta}_{LR}$ is the thermo-field double (TFD) state at temperature $\beta$. The left and right sides of the thermo-field double will be denoted by $L$ and $R$, respectively.
When the Hamiltonian is chaotic, this code of course just realizes the Hayden-Preskill thought experiment of \cite{Hayden:2007wr}, where in this case Alice's diary is just a single qubit's worth of information. 

In this work, we attempt to compute the minimal number of physical qubits (operators on the physical Hilbert space) needed to reconstruct the logical qubit (operators on the code subspace). This notion was actually introduced in the work of Pastawksi \& Preskill under the well-chosen name of \emph{price}, since it is the minimum cost of reconstructing the logical qubit \cite{Pastawski:2017wm}. It is dual to another notion, called the \emph{distance}, which is the smallest number of qubits such that the code can recover from erasures of those qubits. For general codes the distance and price are different, but for codes with complementary recovery, the two notions are the same. Indeed, the codes we consider in the context of SYK will have complementary recovery.

In this work, we relate the price to a conceptually similar concept of operator size, which was explored in \cite{Roberts:2018vv, Qi:2018wg, Susskind:2019ue}. One can expand the Heisenberg evolved operator $\psi(T_R)$ into a complete, orthonormal set of operators
\begin{align}
    \psi(T_R) = \sum c_P P,
\end{align}
where the operators $P$ are operators of definite size $|P|$ and $c_P$ is the size wavefunction of $\psi$. 

The authors in \cite{Brown:2019aa} discussed a phenomenon called size-winding, meaning that $c_P = r_P e^{i \alpha |P|}$ where $r_P \in \mathbb{R}$ and the phase $\alpha$ is independent of $P$. It is expected that SYK is size-winding at least at low temperatures. Assuming the operator is perfectly size-winding, in Sec. \ref{sec:size} we will argue for a simple bound relating the operator size to the price, which comes from thinking about reconstructing the operator $\psi(T_R)$ via quantum teleportation. The bound says that the number of qubits, $K$, of the total number, $N$, that one needs to add from the right to the left in order to successfully perform the teleportation by size protocol of \cite{Brown:2019aa} obeys the constraint
\begin{align}\label{eqn:priceboundintro}
K \gtrsim K_0 = \frac{\beta \braket{H_R - H_L}_{\psi}}{S(T_R)},
\end{align}
where $\braket{H_R - H_L}_{\psi}$ is the energy of the state $\psi(T_R) \ket{\beta}$ and $S(T_R)$ is the average size of the operator $\psi(T_R) \rho_R^{1/2}$ as a function of time.\footnote{To argue for this bound, we will also assume that the size is in a regime of exponential growth in time.} The bound in eq. \eqref{eqn:priceboundintro} implies a lower bound on the price since one needs at least $N-K_*$ qubits on the right to prevent the ability of the complement region to successfully teleport $\psi_R(T_R)$ from the right to the left.

\begin{figure}
\centering
\includegraphics[width=.5\linewidth]{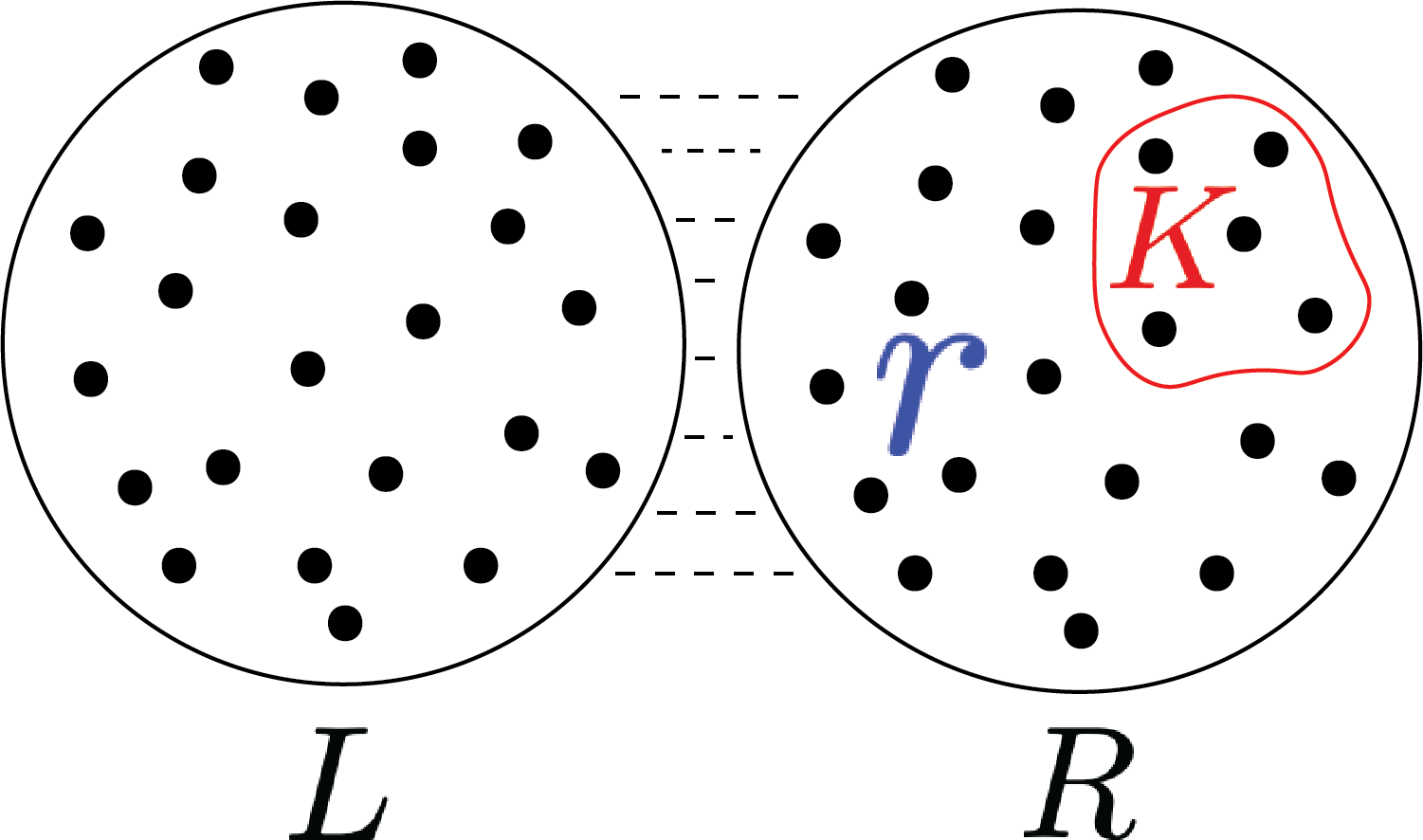}
\caption{We imagine that we have a system of qubits prepared in their thermo-field double state, whose entanglement is represented by the dashed lines. We will ask the question of what is the largest number of qubits, $K$, from the right that we can give to the left such that we can just reconstruct an excitation on $r = \overline{LK}$.}
\label{fig:LKrdefinition}
\end{figure}
We will then compute the price of a single, time-evolved fermion in SYK. We will show that the price approximately saturates the bound implied by eq. \eqref{eqn:priceboundintro}.\footnote{Really we will show that $N$ minus the price parametrically saturates the bound in eq. \eqref{eqn:priceboundintro}, but we will refer to this as saturation of the price bound.} To compute the price, we need to compute the size of a subset $r \subset R$ such that there just barely exists Hermitian operators supported on $r$ that can reproduce the action of all logical operators on $\mathcal{H}_{code}$, defined in eq. \eqref{eqn:Hcode}. See fig. \ref{fig:LKrdefinition} for our region conventions. A similar calculation in more general systems was recently done in \cite{Chandrasekaran:2021wx}, which we follow. In \cite{Chandrasekaran:2021wx} it was argued that the price is computable from the modular flowed correlation function
\begin{align}\label{eqn:modflowcorr}
    \braket{\psi_L(T_L) \Delta_r^{is} \psi_R(T_R)}_{\beta},
\end{align}
where $\Delta_r = \rho_r \otimes \rho_{\overline{r}}$ with $\overline{r}$ the complement of $r$. We will often denote $\overline{r} = LK$ where $K$ is the complement of $r$ within $R$.

When this correlation function has a particular $s$-dependence of the form
\begin{align}\label{eqn:maxmodchaos}
     \braket{\psi_L(T_L) \Delta_r^{is} \psi_R(T_R)}_{\beta} \sim \frac{1}{\left(\cosh(\pi (T_L+T_R+s)) - c_Q \frac{K}{N}\sinh(\pi s)e^{\pi (T_L-T_R)}\right)^{2\Delta}}
\end{align}
then \cite{Chandrasekaran:2021wx} showed that there is a sharp value for the value of $K$, which we denote $K_*$, for which the operator $\psi(T_R)$ is reconstructable from the remaining $N-K_* = |r_*|$ fermions and no fewer. The value can be computed to be \cite{Chandrasekaran:2021wx}
\begin{align}
    K_* = N e^{-2\pi T_R}/c_Q,
\end{align}
which leads to a formula for the price of $\psi(T_R)$:
\begin{align}\label{eqn:priceformula}
    \text{price}_{\beta}(T_R) = N\left(1- e^{-2\pi T_R}/c_Q + \mathcal{O}(e^{-4\pi T_R})\right),
\end{align}
where we have taken the large $T_R$ limit. The main result of this paper will be that for $r$ a subset of SYK fermions of size $N-K$, the modular flowed correlator in eq. \eqref{eqn:modflowcorr} has the form given in eq. \eqref{eqn:maxmodchaos}, at least to leading order in a small $K/N$ expansion, with $K, N \gg 1$. We comment on higher orders in $K/N$ in Sec. \ref{sec:discussion}.

Following the work of \cite{Boer:2019td} and for reasons we will explain below, we will refer to the form of the correlator in eq. \eqref{eqn:maxmodchaos} as exhibiting \emph{maximal modular chaos}. It was argued in \cite{Chandrasekaran:2021wx} that in more conventional, higher dimensional situations in AdS/CFT, maximal modular chaos comes about because the modular operator $\Delta_r$ generates boosts about a bulk quantum extremal surface. Thus, the correlator in eq. \eqref{eqn:maxmodchaos} can be used to make a sharp prediction for the position of the QES in the bulk dual. It is natural that the physics of the QES appears in the calculation of the price since, vis a vis the work of \cite{Harlow:2016vwg}, we know that quantum error correction in holographic systems is inextricably linked to entanglement wedge reconstruction in the bulk dual. 
\begin{figure}
\centering
\includegraphics[width=.5\linewidth]{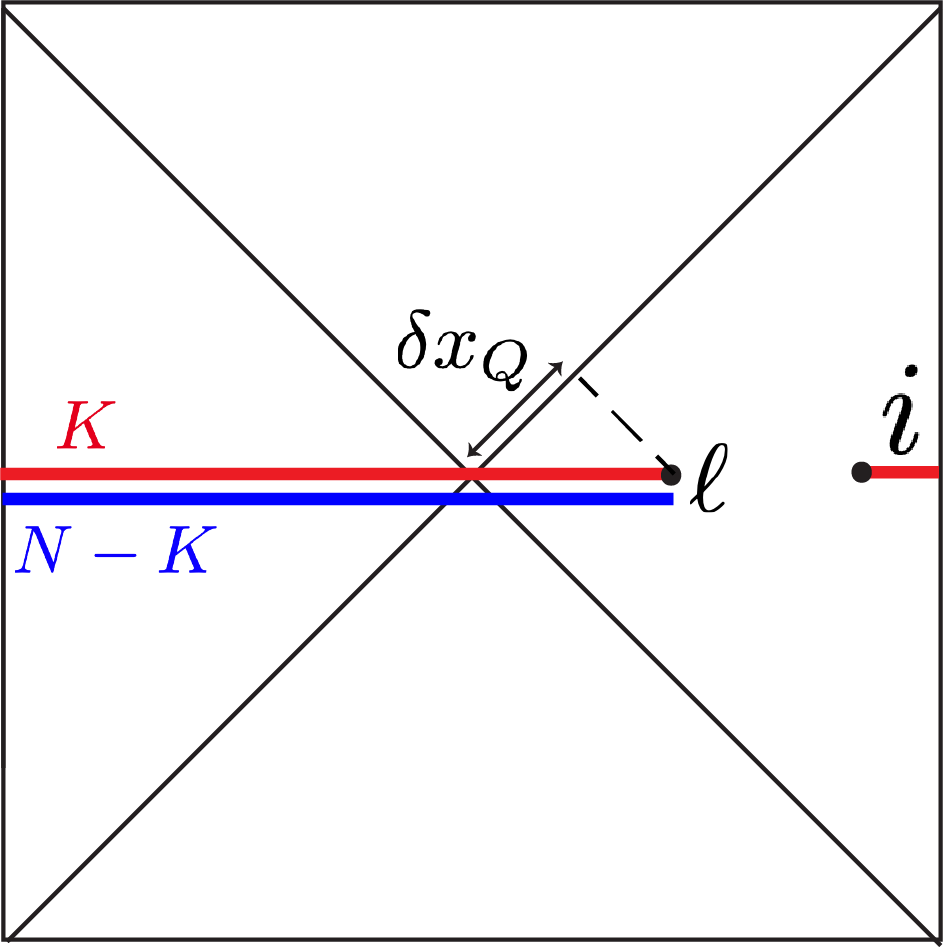}
\caption{We illustrate the putative bulk picture that we test in Sec. \ref{sec:bulkpicture}. We imagine a bulk with $N$ free fermions of mass set by the boundary dimensions of the SYK fermions in the IR limit. We then imagine finding a QES in the bulk associated to $L \cup K$ by fixing a small interval $\boldsymbol{i}$, associated with the $K$ fermions on the right, and then extremizing the generalized entropy for $\ell \cup i$ where only the $K$ fields are included in the entropy for $\ell \cup i$. The remaining $N-K$ fields only contribute to the entropy in $\ell$.}
\label{fig:bulkpicture}
\end{figure}

It is then natural to ask whether there is a quantum extremal surface associated to a subset of spins in SYK. In this work, we will consider a simple bulk dual of SYK which consists of $N$ free fermions with mass set by the dimension of the boundary fermions in the conformal limit. We will couple the fermions perturbatively to JT gravity and then attempt to compute the position of a QES associated to a subset of boundary fermions. Our guess for the bulk QES will be to take a small, fixed interval, $\boldsymbol{i}$, near the boundary of AdS. We then extremize the generalized entropy
\begin{align}
S_{\text{gen}} (\ell I) = \phi(x) + (N-K) S_{\text{bulk}}(\ell) + K S_{\text{bulk}}(\ell I)
\end{align}
where $\ell$ is a region which stretches all the way from the left boundary to some point $x$. As discussed in Sec. \ref{sec:bulkpicture}, we find general, parametric agreement with the position predicted by the correlator in eq. \eqref{eqn:maxmodchaos}.

An interesting consequence of our results is that in the large $N$ limit, the operator algebras associated to subsets of fermions turn into Type III${_1}$ von Neumann algebras. More precisely, the modular operators $\Delta_{LK}$ and $\Delta_L$ satisfy the inclusion
\begin{align}
    \Delta_{LK}^{is}\Delta_L^{-is} = \text{exp}\left(i(e^{-2\pi s}-1)\delta x^+_Q P_+ \right),
\end{align}
where $\delta x^+_Q$ is the position of the QES in the bulk. This relation is known as a half-sided modular inclusion. From the perspective of the emergent Type III$_1$ boundary algebra, $P_+$ is just an abstract operator in the inclusion algebra, with no geometric interpretation in and of itself. However, $P_+$ corresponds in the bulk to the null translation generator discussed in \cite{LMZ}. In this sense, null translations in the bulk emerge from boundary half-sided modular inclusions of subsets of spins. This is a concrete manifestation of bulk emergence from non-spatial degrees of freedom. It is also analogous to the results of \cite{Leutheusser:2021aa, Witten:2021aa}, which find an emergent Type III$_1$ von Neumann algebra in the large $N$ limit of AdS/CFT, for time bands of a given side of the TFD state.  

We now give a brief outline. In \textbf{Sec.} \ref{sec:size}, we define more carefully the notion of price and compare it to operator size, arguing for the bound in eq. \eqref{eqn:priceboundintro}. We review the argument of \cite{Chandrasekaran:2021wx} that price can be computed via modular flowed correlators like that in eq. \eqref{eqn:modflowcorr}. In \textbf{Sec.} \ref{sec:setup}, we briefly review the SYK model and discuss more carefully how we compute price in this model. We discuss the replica trick that we use to compute the modular flowed correlator for a subset of spins. In \textbf{Sec.} \ref{sec:SDsolution}, we then solve the large $N$ Schwinger-Dyson equations of SYK with boundary conditions set by this replica trick. In \textbf{Sec.} \ref{sec:analyticcontinuation}, we perform the analytic continuation necessary to implement this replica trick and prove that, at low energies, the SYK model exhibits maximal modular chaos for a subset of fermions of size $N-K$ with $K \ll N$. Finally, in \textbf{Sec.} \ref{sec:bulkpicture}, we compare with our proposed bulk picture in terms of a QES. In \textbf{Sec.} \ref{sec:discussion}, we end with a brief discussion of our results and future directions. In the appendices, we collect useful formulae and do some supplementary calculations.

\section{Price of error correcting codes and operator size}\label{sec:size}
In this section we define more precisely the price of a quantum error correcting code, and discuss its relation to the notion of operator size considered in \cite{Roberts:2018vv, Qi:2018wg, Susskind:2019ue}. We first present a general formulation of these concepts, and then connect them explicitly in SYK. For our purposes it will be convenient to do this in the language of operator algebra quantum error correction, whose salient aspects we now briefly summarize, following \cite{Pastawski:2017wm, Harlow:2016vwg}.\footnote{While our discussion is framed in terms of finite dimensional systems of qubits, the definitions carry over straightforwardly in the large $N$ limit simply by computing everything at finite $N$ and keeping track of the scaling as we take $N \rightarrow \infty$.} 

Consider a von Neumann algebra $\mathcal{A}$ acting on a Hilbert space $\mathcal{H}_{\text{code}}$, which is a code subspace of some physical Hilbert space $\mathcal{H}$. We can decompose the physical Hilbert space into a tensor product $\mathcal{H} = \mathcal{H}_A \otimes \mathcal{H}_{\bar{A}}$. The algebra $\mathcal{A}$ consists of the logical operators which map $\mathcal{H}_{\text{code}}$ onto itself. We will mainly be interested in correctability of erasure quantum channels. In particular, the algebra $\mathcal{A}$ is protected against erasure of $\bar{A}$ if there exists a recovery channel, $\mathcal{R}$, which yields the identity channel when composed with the erasure channel, $\mathcal{R} \circ \text{Tr}_{\bar{A}}=1$. 

Importantly, one can show \cite{Pastawski:2017wm} that erasure of $\bar{A}$ is correctable if for each operator $O$ in $\mathcal{A}$, there exists a logically equivalent operator $O_{A}$ supported on $A$, meaning
\begin{align}
    O \ket{\psi} = O_{A} \ket{\psi},\ \forall \ket{\psi} \in \mathcal{H}_{\text{code}}.
\end{align}
We then say that $\mathcal{A}$ can be \emph{reconstructed} on $A$. 

We now come to the definition of the price of the code. The \emph{price}, $p(\mathcal{A})$, is the size of the smallest region $A$ such that for every operator in $\mathcal{A}$ there is a logically equivalent operator supported on $A$. In other words, the price tells us how small we can make $A$ before the logical algebra is no longer reconstructable on $A$. In equations, the price is defined as
\begin{align}
    p(\mathcal{A}) = \min_A \lbrace |A|: \forall \mathcal{O} \in \mathcal{A},\ \mathcal{O}\text{ is reconstructable in A}\rbrace,
\end{align}
where $|A|$ denotes the number of physical qubits in $A$.

\subsection{The price of an operator}\label{sec:priceofop}
We can use price to motivate a notion of operator size as follows. Consider a Hilbert space $\mathcal{H}$ which corresponds to a set of physical qubits. Given an operator $O_{\text{phys}}$ acting on the physical Hilbert space $\mathcal{H}_{\text{physical}}$, along with some reference state $\ket{\phi} \in \mathcal{H}_{\text{physical}}$ such that $\braket{O_{\text{physical}}}_{\phi}=0$, we can construct the algebra $\mathcal{A}^{\psi}_O \subset \mathcal{B}(\mathcal{H}_{\text{physical}})$ generated by\footnote{Without loss of generality, we can assume $O$ has zero expectation value in $\ket{\phi}$. Otherwise, we subtract from $O$ its mean value in $\ket{\phi}$.}
\begin{align}\label{eqn:Aodef}
   O_{\text{phys}}\ket{\phi}\bra{\phi}
\end{align}
and its hermitian conjugate. This is isomorphic to the algebra of a single qubit. We then define the \emph{price of $O$ with respect to $\ket{\phi}$} which is just
\begin{align}
    p_{\phi}(O) \equiv p(\mathcal{A}_O^{\phi}).
\end{align}
As mentioned in the introduction, the price is in general dual to the distance of the code, which is the size of the smallest region $A$ that is not correctable given $\mathcal{A}$. In the special case of codes with complementary recovery, the price is equal to the distance \cite{Pastawski:2017wm}. 

\subsection{Relation to operator size}\label{sec:pricesize}
We now analyze the relationship between the price of an operator and its size. It seems natural that the two concepts should be related since they are both intuitively a measure of how difficult it is to reconstruct the action of an operator.

A simple lower bound on the price of an operator can be derived by relating bulk reconstruction to quantum teleportation \cite{GJW, Brown:2019aa, Nezami:2021td}. This can be found by lower bounding the number of fermions $K$ on the right that one needs to give to the left in order to successfully teleport the operator. It is not hard to see that if, for $K \geq K_0$, the action of an operator $O$ on the thermofield double can be teleported, then the price of that operator obeys $p_{\beta}(O) \geq N - K_0$. When 

By ``teleporting" an operator, what we mean is that there exists a simple bilocal unitary $e^{igV}$ with $V  \sim \sum \phi_L \phi_K$ such that $e^{igV} O_R\ket{\beta} = O_L\ket{\beta}$ where $O_L$ is the mirror of $O_R$. Here we are imagining that $\phi_L$ and $\phi_K$ are simple operators. The authors of \cite{Brown:2019aa, Nezami:2021td, Schuster:2021uc} argued that the phenomenon of size winding is intimately related to the ability to perform quantum teleportation. 

As a toy model for our problem, we can imagine a system of $N$ qubits, acted on by the Pauli matrices at each site. Following \cite{Schuster:2021uc}, we can define the \emph{$K$-size operator} as
\begin{align}
    V_K = \frac{1}{K} \sum_{i=1}^K \left(1-\frac{1}{4}\sum_{P^i} P_L^i (P_R^i)^*\right)
\end{align}
where $P_{L(R)}^i$ is the left (right) Pauli acting on qubit $i$. The sum runs over $P^i =1, X,\,Y,\,Z$. Here we imagine that the background is the thermofield double of two copies of the spin system. This operator measures the \emph{$K$-size} of Pauli $P$ i.e. the number of non-identity Pauli operators supported on the $K$ subsystem that have been acted on the maximally entangled state (infinite temperature state). In particular, in the case where $K=N$, 
\begin{align}
    V_N \ket{P} = \frac{|P|}{N}\ket{P},
\end{align} 
where $|P|$ is the size of $P$. 

A general, $N$-qubit operator can be expanded into a complete basis of definite size operators as
\begin{align}
    O = \sum_P c_P P,
\end{align}
where $c_P$ is the size wavefunction of $O$. Systems with a holographic dual are expected to exhibit a phenomenon called \emph{size-winding} \cite{Brown:2019aa}, where the wavefunction $c_P$ has a phase which depends linearly on the size of the Pauli $P$. Explicitly,
\begin{align}
    c_P = e^{i \alpha |P|} r_P
\end{align}
where $\alpha \in \mathbb{R}$ is the same for all $P$ and $r_P \in \mathbb{R}$. Given a Hermitian operator $O_R(T_R)$, which is the time evolved version of a few-qubit operator, the size-winding ansatz as applied to the non-Hermitian operator $O_R(T_R) \rho_R^{1/2}$ gives
\begin{align}\label{eqn:opsizeexp}
    O_R(T_R)\ket{\beta} = O_R(T_R) \rho_R^{1/2} \ket{0} = \sum_P e^{i\alpha |P|}r_P \ket{P},
\end{align}
where $\ket{P} = P\ket{0}$, and $\ket{0}$ is the maximally entangled $\beta = 0$ state. Here $c_P$ is the wavefunction of the operator $O_R \rho^{1/2}_R$. Then the average size of $O_R(T_R) \rho_R^{1/2}$ is  
\begin{align}
    S(O_R(T_R)\rho^{1/2}) = \sum_P r_p^2 |P|,
\end{align}
where we have assumed that the $c_P$'s are normalized. The mirror operator $O_L(T_L)$ acts on the thermofield double as
\begin{align}\label{eqn:mirroropsizeexp}
    (O_L(-T_R))^* \ket{\beta} =\rho_R^{1/2} O_R^*(T_R) \ket{0} = \sum_P e^{-i\alpha |P|}r_P \ket{P}.
\end{align}
From eqs. \eqref{eqn:opsizeexp} and \eqref{eqn:mirroropsizeexp} it is clear that when $O_R(T_R) \rho_R^{1/2}$ displays size winding, teleportation can be accomplished just by acting with the operator $e^{igV_N}$ and adjusting the coefficient $g$ to be
\begin{align}\label{eqn:galpha}
    \frac{g}{N} = -2\alpha.
\end{align}
Note that here we will assume that the operator $O_R \rho_R^{1/2}$ displays what the authors of \cite{Brown:2019aa} call \emph{detailed winding}, which means that the coefficient $\alpha$ is a constant, independent of the individual Pauli string.

Given detailed winding, one can write down a differential equation for $\alpha(T_R)$ in terms of the size of $O_R(T_R) \rho_R^{1/2}$. Differentiating the state $O_R(T_R)\ket{\beta}$ with respect to $T_R$, it is not hard to see that
\begin{align}
   S(T_R) \frac{d}{dT} \alpha(T_R) = \braket{H_R - H_L}_{O_R(T_R)\ket{\beta}}
\end{align}
where $\braket{H_R - H_L}_{O_R(T_R)\ket{\beta}}$ is the boost energy of the state $O_R(T_R)\ket{\beta}$. To derive this, we also used that
\begin{align}
  \sum_P r_P(T) \frac{d}{dT}r_P(T)= \frac{1}{2} \frac{d}{dT} \sum_P r_P^2  = \frac{1}{2}\frac{d}{dT} ||O(T) \ket{\beta}||^2=0.
\end{align}

Since $\braket{H_R- H_L}_{O\ket{\beta}}$ is independent of $T_R$, we see that the dynamics of $\alpha(T_R)$ is entirely fixed by the size. If the size is growing exponentially as $S(t) \sim e^{2\pi t/\beta}$, this differential equation can be solved easily and we find
\begin{align}\label{eqn:alphaeqn}
    \alpha(T_R) = -\frac{\beta}{2\pi} \frac{\braket{H}_{O\ket{\beta}}}{S(T_R)}
\end{align}
where we have assumed that $\alpha \to 0$ at large $T_R$.\footnote{This assumption is reasonable from the bulk dual point of view. If we drop this assumption, eq. \eqref{eqn:alphaeqn} will get modified by boundary terms for $\alpha$ which will need to be determined by some other method.} Thus we see from eq. \eqref{eqn:galpha} that as $T_R$ increases, we require smaller and smaller $g$ in order to successfully teleport the operator.

It is then clear that given access to only a subset $K$ on the left and right, and assuming that the operator $O\rho^{1/2}$ exhibits detailed winding, it is sufficient for quantum teleportation that the following identity holds:\footnote{In SYK, we expect this identity to hold for operators that have been sufficiently scrambled and for $K \gg 1$.}
\begin{align}\label{eqn:KsizeNsize}
    e^{ig K V_K}O_R(T_R)\ket{\beta} \approx e^{ig K V_N} O_R(T_R) \ket{\beta},
\end{align}
for a range of $g K$ such that 
\begin{align}
    \frac{gK}{N} \geq -\alpha(T_R) = \frac{\beta}{2\pi} \frac{\braket{H}_{O \ket{\beta}}}{S(T_R)}.
\end{align}
In order to claim a reconstruction of $O_R$ on $LK$ that is Hermitian, we also would like for the operator $e^{igKV_K}$ to approximately preserve the thermofield double up to an overall phase:
\begin{align}
    e^{igKV_K}\ket{\beta} \approx e^{igK \braket{V_K}_{\beta}} \ket{\beta}
\end{align}
which, in a large $N$ model like SYK, tells us that $g$ needs to be relatively small, $g \lesssim 1$ (i.e. not scaling positively with any large parameters in the problem). This can be seen by demanding $\braket{\beta| e^{igK V_K} |\beta} \approx e^{igK \braket{V_K}_{\beta}}$ and computing the left hand side by expanding the exponential and using large $N$ Wick contractions. Given $g \geq 1$, we get a bound on $K$ without $g$, which is
\begin{align}\label{eqn:Kbound}
    K \geq \frac{N \beta}{2\pi} \frac{\braket{H}_{O \ket{\beta}}}{S(T_R)}.
\end{align}
This tells us how big $K$ needs to be in order to successfully teleport the operator $O_R(T_R)$, assuming that $O_R(T_R)$ exhibits detailed size-winding. From the discussion earlier, this leads to a bound on the price with respect to the thermofield double of $O_R(T_R)$, which is
\begin{align}\label{eqn:priceboundsize}
    p_{\beta}(O_R(T_R)) \gtrsim N- K_0(T_R)
\end{align}
where $K_0(T_R) = \frac{N \beta}{2\pi} \frac{\braket{H}_{O \ket{\beta}}}{S(T_R)}$.

One main result of this work is that the price in SYK at large $\beta \mJ$ parametrically saturates this bound, as pointed out in the introduction, eq. \eqref{eqn:priceformula}. Namely, $N -p_{\beta}(O_R(T_R)) \sim K_0(T_R)$. This translates to the statement that there is no parametrically better way of reconstructing the operator $O_R(T_R)$ on $LK$ than quantum teleportation, at least to leading non-trivial order in $K/N$. Similar statements have been discussed in more general models other than SYK \cite{Levine:2020aa,Chandrasekaran:2021wx}. As we will discuss, the parametric saturation of the bound in eq. \eqref{eqn:priceboundsize} is because of the special form of the modular Hamiltonian for $LK$ in SYK at large $\beta \mJ$ and small $K/N$. In particular, the modular Hamiltonian for $LK$ to leading order in small $K/N$ and at large $\beta \mJ$ is approximately proportional to the size operator.

\subsection{Computing the price $p_{\beta}(O)$}\label{sec:pricecalc}
To compute the price, we need to know for fixed $K$ at what time $T_R$ the insertion of $\psi(T_R)$ on the TFD stops being reconstructable from $r$ (or becomes reconstructable on $\bar{r} = LK$). As discussed in Sec. \ref{sec:priceofop}, we can determine when the algebra $\mA_O^{\beta}$ defined around eq. \eqref{eqn:Aodef} is reconstructable on $r$ or $LK$ by asking when the qubit operators 
\begin{align}
    \sigma_X \equiv O\ket{\beta} \bra{\beta} + h.c.,\ \sigma_Y \equiv i O\ket{\beta} \bra{\beta} + h.c.
\end{align} 
are reconstructable on $r$ or $LK$. This calculation was done in \cite{Chandrasekaran:2021wx} where it was argued that one can calculate whether $\mA_O^{\beta}$ is reconstructable on $LK$ by calculating 
\begin{align}\label{eqn:fidelity}
    F\left(\psi^{X,Y}_{\lambda}\big\lvert\beta; LK\right) =  \underset{{U_{r}}}{\text{sup}} \Big\lvert \Big\langle \psi^{X,Y}_{\lambda}\Big\lvert U_{r}\Big\lvert \beta\Big\rangle \Big\lvert^{2},
\end{align}
where 
\begin{align}
    |\psi^{X,Y}_{\lambda}\rangle = e^{i\lambda \sigma_{X,Y}}|\beta\rangle, ~\lambda \in \mathbb{R}. 
\end{align}
Actually \cite{Chandrasekaran:2021wx} argued that to find when $\sigma_{X,Y}$ are reconstructable on $r$ one only needs to know the fidelity perturbatively in $\lambda$. Expanding the fidelity, one has
\begin{align}\label{eqn:susceptibility}
    F\left(\psi^{X,Y}_{\lambda}\big\lvert\beta; LK\right) = 1 - \lambda^2 \chi(\psi^{X,Y}|\beta; LK) + \mathcal{O}(\lambda^3) 
\end{align}
where $\chi(\psi^{X,Y}|\beta; LK)$ is called the fidelity susceptibility. We are therefore looking for the smallest size of $r$ for which the susceptibility $\chi(\psi^{X,Y}|\beta;LK)$ is close to zero.\footnote{Strictly speaking, to show that the Pauli matrices obey the correct algebra, we need to know about the order $\lambda^4$ term, as well. However, based on properties of modular flowed correlators, it was conjectured in \cite{Chandrasekaran:2021wx} that it should display the same qualitative behavior as the order $\lambda^2$.}

Using the work of \cite{May:2018ti}, it was shown in \cite{Chandrasekaran:2021wx} that the essential ingredient to compute the fidelity susceptibility in eq. \eqref{eqn:susceptibility} is just the modular flowed correlator discussed in eq. \eqref{eqn:modflowcorr}. In particular, the susceptibility is directly related to the correlation function
\begin{align}\label{eqn:modflowcorr2}
    \braket{\psi_R(T_R-i\delta) \Delta_r^{is}\psi_R(T_R+i\delta)}_{\beta},
\end{align}
where $\delta$ is the smearing scale defining the state. When this correlator saturates maximal modular chaos and takes the form in eq. \eqref{eqn:maxmodchaos},
\begin{align}\label{eqn:maxmodchaos2}
     \braket{\psi_L(T_L) \Delta_r^{is} \psi_R(T_R)}_{\beta} \sim \frac{1}{\left(\cosh(\pi (T_L+T_R+s)) - c_Q \frac{K}{N}\sinh(\pi s)e^{\pi (T_L-T_R)}\right)^{2\Delta}},
\end{align}
the fidelity sharply transitions in $T_R$ as we make it more negative. In particular, when $\delta$ is small, at a time $T_R(N,K)$ defined by
\begin{align}\label{eqn:TRKN}
e^{2\pi T_R(N,K)} = \delta x_Q = c_Q \frac{K}{N},
\end{align}
the susceptibility rapidly transitions from being zero to order $1-\delta$ over a time-scale of order $\delta$. Here $c_Q$ is the parameter that appears in eq. \eqref{eqn:maxmodchaos}. This sharp transition is related to the appearance of a quantum extremal surface in the bulk at location $x^{\pm} = \delta x_Q$ where $x^{\pm}$ are Kruskal coordinates near the horizon. Very similar calculations were done in \cite{Suzuki:2019uw, Kusuki:2019uh} in the context of higher dimensional CFTs, although the authors were more interested in finding a classical extremal surface whereas in our calculations we will often be most interested in the quantum corrections.

Utilizing these results, we can thus compute the price just by computing the modular flowed correlator in eq. \eqref{eqn:modflowcorr2}, matching its form to that in eq. \eqref{eqn:maxmodchaos2} and extracting $\delta x_Q$. The price is then given by solving \eqref{eqn:TRKN} for $K$ and then setting
\begin{align}
    p_{\beta}(\psi_R(T_R)) = N-K(T_R,N)= N(1-c_Q e^{2\pi T_R}).
\end{align}
In principle, we could do this in any system that displays maximal modular chaos. A particular interesting system will be the SYK model, which we turn to now.

\section{SYK and modular flow}\label{sec:setup}
We now turn to computing the price of an operator in a concrete model: the Sachdev-Ye-Kitaev (SYK) model. We start by briefly reviewing relevant aspects of the SYK model. As is well known, the SYK model consists of $N$ Majorana fermions whose dynamics is given by the Hamiltonian 
\begin{align}
    H = (i)^{q/2}\sum_{1 \leq i_{1} \leq i_{2} \leq...\leq i_{q} \leq N}j_{i_{1}i_{2}\ldots i_{q}}\psi_{i_1}\psi_{i_2}\ldots \psi_{i_q},
\end{align}
where $q$ is an even number, and the coefficients $j_{i_1 i_2 \ldots i_q}$ are drawn from an ensemble of Gaussian random variables:
\begin{align}
    \langle j_{i_1 i_2 \ldots i_q}^2 \rangle = \frac{J^2 (q - 1)!}{N^{q-1}}.
\end{align}

By integrating out the fermions, the partition function of the theory can be written in terms of a pair of bilocal fields $G(\tau,\tau'),~\Sigma(\tau,\tau')$ as\footnote{Here and throughout the paper, we always mean by $Z$ the ensemble averaged partition function.} \cite{Maldacena:2016vs} 
\begin{align}
    Z &= \int DG D\Sigma e^{-N S[G,\Sigma]},\\
    S[G,\Sigma] &= -\frac{1}{2}\text{log}~\text{det}\left(\partial_{\tau}\delta(\tau-\tau') - \Sigma\right) + \frac{1}{2}\int d\tau \int d\tau' \left(\Sigma G - \frac{1}{q}J^2 G^q \right), 
\end{align}
where $\tau$ is the Euclidean time coordinate, and $G(\tau, \tau') = \frac{1}{N}\sum_i \langle \psi_i(\tau)\psi_i(\tau')\rangle$ is the two-point function. It will be often convenient to treat $G,\ \Sigma$ as matrices in $\tau$ and $\tau'$ with a matrix product 
\begin{align}
    M \star N \equiv \int d\tau'' M(\tau, \tau'')N(\tau'', \tau').
\end{align}
At large $N$ we can do a saddle-point approximation, in which case the action is evaluated on the classical solutions for $G$ and $\Sigma$, which are given by the Schwinger-Dyson equations, 
\begin{align}
    G(\tau,\tau') &= (\partial_{\tau} - \Sigma)^{-1}\\
    \Sigma(\tau,\tau') &= J^2 \left(G(\tau,\tau')\right)^{q-1}. 
\end{align}

Following the ideas in Sec. \ref{sec:size}, we will consider two copies of SYK entangled in the thermo-field double state (TFD) with respect to the SYK Hamiltonian. We will denote the two copies by $L$ and $R$. We then divide $R$ into two parts: a set of fermions $K$ and the remaining fermions $N - K$. We will take $K$ to be even so that we can associate a Hilbert space to the $K$ fermions. In this work, we will only go to first order in the limit $K/N \ll 1$.\footnote{Note that we will overload the symbol $K$ to mean both the number of fermions in $K$ and the set of fermions labeled by $K$. Context should elucidate which we mean.} We will denote the set of the remaining $N-K$ fermions in $R$ by $r$.

We prepare the $LR$ system in the TFD state, which can be compactly written as \cite{Qi:2018wg}
\begin{align}
|\beta\rangle = Z_{\beta}^{-1/2}e^{-\frac{\beta}{4}(H_L + H_R)}|0\rangle, 
\end{align}
where $\beta$ is the inverse temperature, and $|0\rangle$ is the maximally-entangled, fermion ground state which obeys 
\begin{align}
(\psi^j_L + i \psi^j_R)|0\rangle = 0,\ \forall j. 
\end{align}
We will act on the TFD state with one of the fermions and keep track of the perturbation as it grows. For convenience, we will take the perturbing fermion, $\psi^j_R$, to be in the set $r$, although the answers do not depend on that choice, as detailed in Appendix \ref{app:Kfermion}. In the end, we will average the modular flowed correlation function over the choice of fermion in $r$. To that aim, consider the following state: 
\begin{align}
|\psi_{\delta}^j\rangle =Z_{\delta}^{-1/2} \Delta_R^{\delta}\psi^j_R|\beta\rangle,
\end{align}
where $\Delta_R = \rho^{-1}_L \otimes \rho_R$ is the full modular operator of the TFD state (recall that for the TFD state, $\Delta_R$ is simply the time evolution operator). We have evolved the insertion with a small amount of Euclidean time, $\delta$, in order to make the state normalizable, with normalization $Z_{\delta}$. 

As we evolve the insertion back in Lorentzian time using $\rho_R$, 
\begin{align}
\psi^j_R(T_R) := \rho_R^{-iT_R}\psi^j_R(0)\rho_R^{iT_R}, 
\end{align}
the perturbation will grow in size as defined in Sec. \ref{sec:size} until, at some time $T^*_R$, the entanglement in the TFD state allows the perturbation to be reconstructed on $L \cup K$ and stops being reconstructable on $r$. Our goal is to find this time $T_R^*$ as a function of $K$. 

As discussed in Sec. \ref{sec:size}, in order to detect when the perturbation is reconstructable from $L \cup K$, we need to compute the modular flowed correlation function 
\begin{align}
    \braket{\Delta_{LK}^{is}}_{\psi_{\delta}} = \frac{1}{N-K}\sum_{i=1}^{N-K} \braket{\psi_{\delta}^i|\Delta_{LK}^{is}|\psi_{\delta}^i}_{\beta},
\end{align}
where again $\Delta_{LK} = \rho_{LK} \otimes \rho_r^{-1}$. Here we have averaged the correlator over the choice of perturbing fermion in $r$, since this choice should only have sub-leading in $1/N$ effects. Modular-flowed correlators of this type were considered in \cite{Balakrishnan:2017aa} and a replica-trick for computing such quantities was discussed in \cite{Faulkner:2018vl}. We will utilize this replica trick here. The trick is to compute
\begin{align}\label{eqn:intnmodflowcorr}
    \frac{Z_{\delta}^{-1}}{N-K}\sum_{i=1}^{N-K}\braket{\psi^i_L(T_L)\left(\rho_{LK}^{p} \otimes \rho_r^{n-1-p}\right)\psi_R^i(T_R)}
\end{align}
at integer $n,\ p$ and then take $n\to 1$ and $p \to is$. To compute this correlator, we think of it as a two point function of fermions in an $n$ copy theory with differing boundary conditions for the $r$ fermions versus the $K$ fermions.

\begin{figure}
\begin{subfigure}{1\linewidth}
\centering
\includegraphics[width=.4\linewidth]{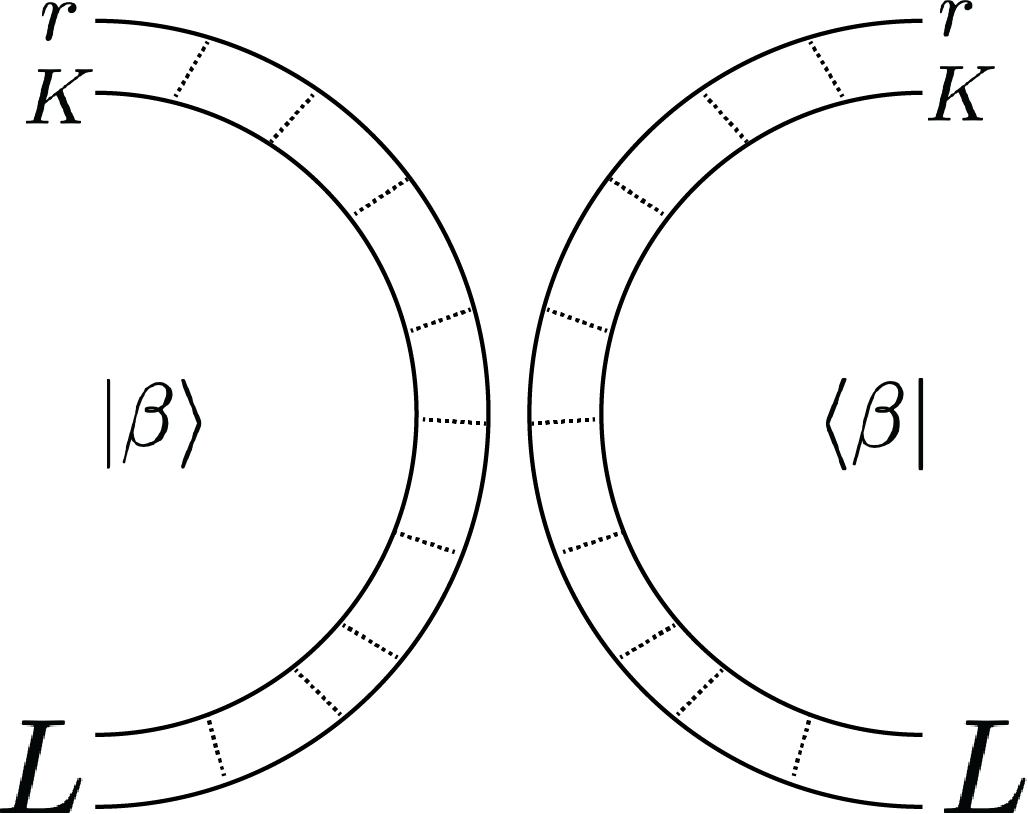}
\caption{}
\label{fig:rhotfd}
\end{subfigure}
\begin{subfigure}{.5\linewidth}
\centering
\includegraphics[width=\linewidth]{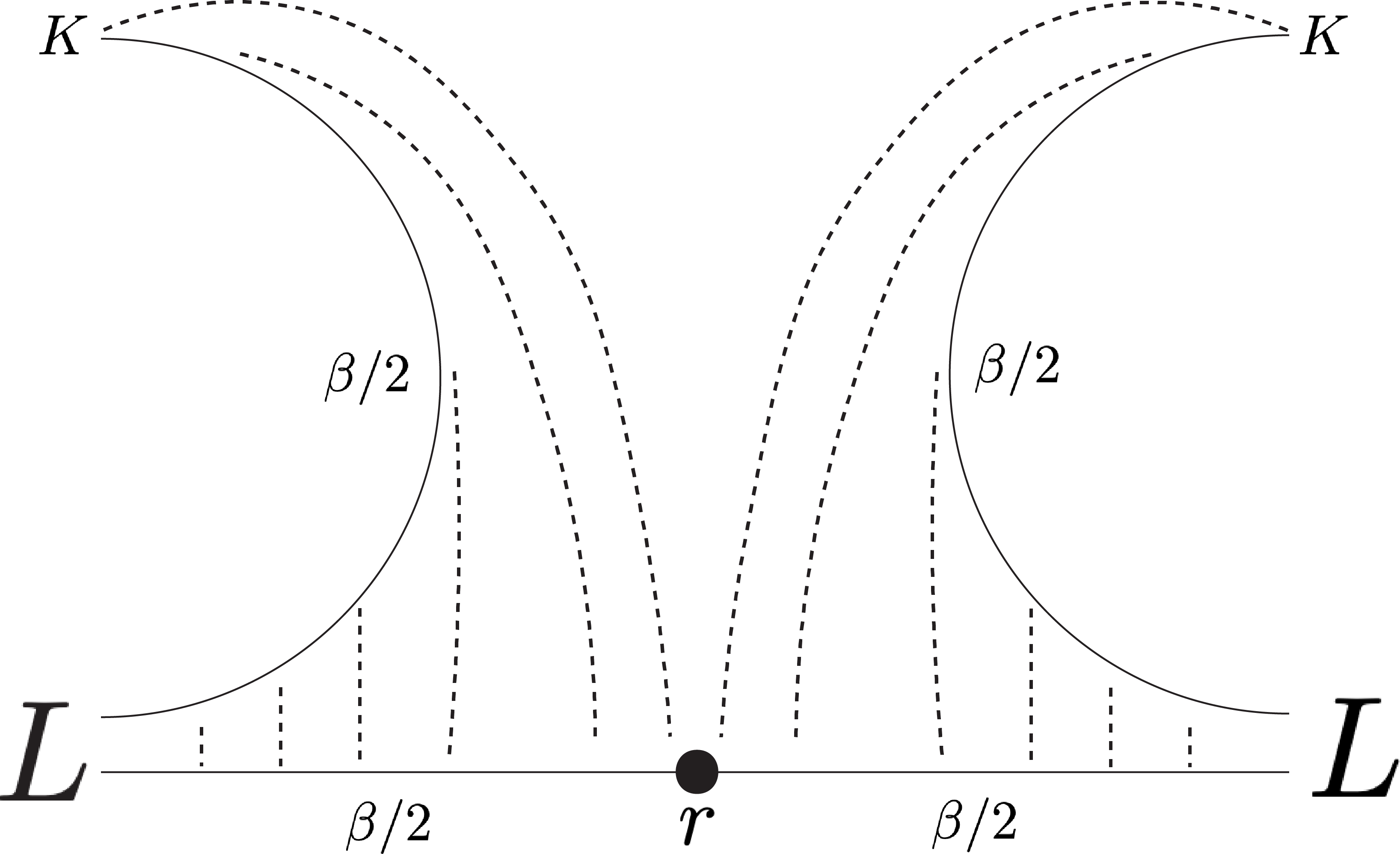}
\caption{}
\label{fig:rhoLK}
\end{subfigure}%
\begin{subfigure}{.5\linewidth}
\centering
\includegraphics[width=\linewidth]{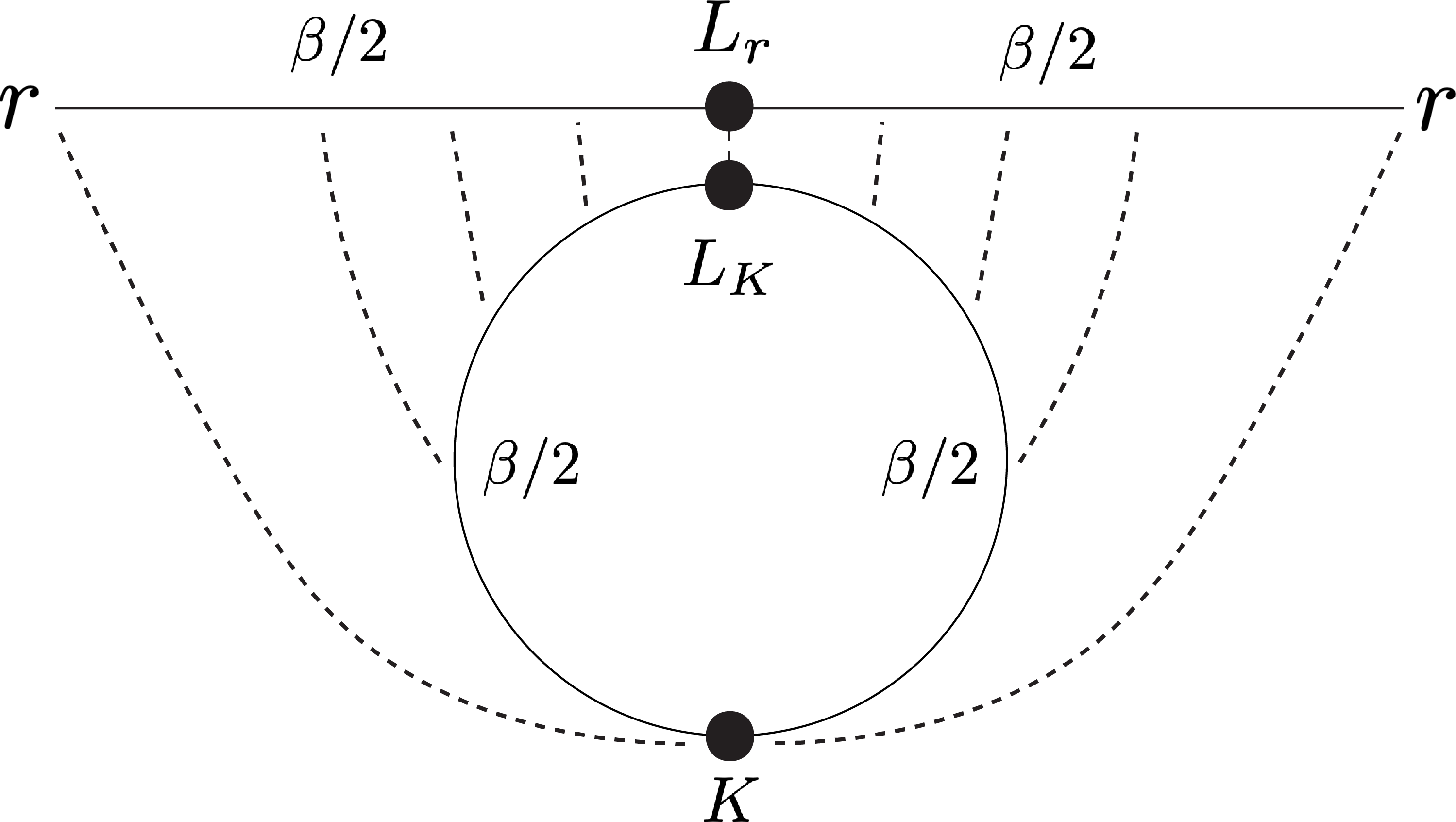}
\caption{}
\label{fig:rhor}
\end{subfigure}
\caption{The top figure illustrates the path integral one must do to prepare the pure state density matrix $\ket{\beta}\bra{\beta}$. The bottom two figures illustrate tracing out subsystems from the TFD to produce a mixed state density matrix. In Fig. \ref{fig:rhoLK}, we have traced out $r$ to produce $\rho_{LK}$, and in Fig. \ref{fig:rhor} we have traced out $LK$ to produce $\rho_r$. The dashed lines remind us that the two ``wires'' (contours) are coupled via Euclidean evolution with the SYK Hamiltonian. Thus, they tell us what parts of each wire have the same Euclidean time. Contours of Euclidean time length $\beta/2$ have been indicated. $L_r$ and $L_K$ refer to the image of $r$ and $K$ under Euclidean evolution by $\beta /2$.}
\label{fig:rhos}
\end{figure}
\begin{figure}
\centering
\includegraphics[width=\linewidth]{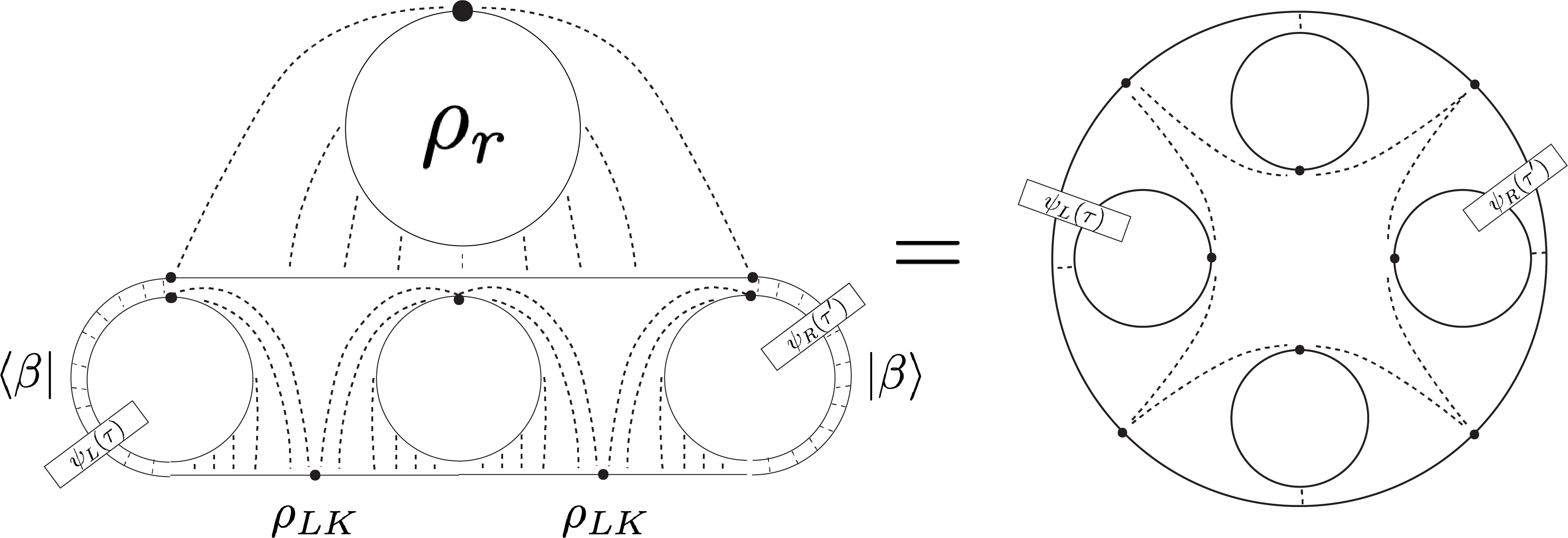}
\caption{On the left we illustrate the correlator $\bra{\beta} \psi_L(\tau) (\rho_{LK}^2 \otimes \rho_r) \psi_R(\tau')\ket{\beta}$. Since only the topologies of the wires and the dashed lines are meaningful, we can write this diagram as on the right, where it looks like a standard replica contour. Black dots connected by dashed lines are at the same Euclidean time.}
\label{fig:correlator}
\end{figure}

It is easiest to see what the best boundary conditions are by working graphically. We can form $\rho_r$ by taking the pure state density matrix and tracing out $LK$: $\rho_r = \text{Tr}_{LK} \ket{\beta}\bra{\beta}$. This is represented in Fig. \ref{fig:rhor}. Similarly, we can form $\rho_{LK}$ by tracing over $r$, which is illustrated in Fig. \ref{fig:rhoLK}. Then we can compute the correlator by forming the diagram shown in Fig. \ref{fig:correlator}. We illustrate the choice $p =2,\ n=4$. From this diagram, we see that the fermions in $K$ have the boundary condition of living on $n$ disconnected $\beta$-length circles while the $r$ fermions live on a single $n\beta$-length circle. Similar diagrams were drawn in \cite{Zhang:2020us}.

To find the $G$-$\Sigma$ equations for this contour, we begin with an $n$ copy theory where we impose boundary conditions in the manner dictated by Fig. \ref{fig:correlator}. The action of the $n$-replica theory is
\begin{align}
S_E = \sum_{k=0}^{n-1} \left(\psi^{(k)}_i \partial_{\tau} \psi^{(k)}_i + \sum_{I} J_I \psi^{(k)}_I\right),
\end{align}
where $I$ is a super-index denoting $i_1,..., i_q$ and $\psi^{(k)}_I = \psi^{(k)}_{i_1}... \psi^{(k)}_{i_q}$ and the superscripts denote a replica index. Note that $J$ is the same for all replicas so that when we integrate over $J$, we get cross replica contributions. 

 Now we introduce the $G,\Sigma$ variables
\begin{align}
&G_r^{(j,k)}(\tau, \tau') = \frac{1}{N-K}\sum_{i=1}^{N-K} \psi^{(j)}_i (\tau) \psi^{(k)}_i(\tau'), \ i \in r, \nonumber \\
&G_K^{(j,k)}(\tau, \tau') = \frac{1}{K}\sum_{i=1}^{K} \psi^{(j)}_i (\tau) \psi^{(k)}_i(\tau'), \ i \in K.
\end{align}
Following the standard steps and integrating out the fermions, the $G,\Sigma$ action becomes
\begin{align}\label{eqn:GSigmaactionreplica}
-S_E/N = &\frac{1}{2} \left( \lambda\text{Tr} \log (\partial_{\tau}\textbf{1} -  \mathbf{\Sigma}_K)  + (1-\lambda)\text{Tr} \log (\partial_{\tau}\mathbf{1} -  \mathbf{\Sigma}_r)\right) \nonumber \\
&-\frac{1}{2}\sum_{j,k=0}^{n-1}\int_0^{\beta} d\tau d\tau' \left( (1-\lambda) \Sigma^{(j,k)}_r(\tau, \tau') G^{(j,k)}_r(\tau, \tau') + \lambda  \Sigma^{(j,k)}_K(\tau, \tau') G^{(j,k)}_K(\tau, \tau')\right. \nonumber \\
&\hspace{2.7cm} \left.- \frac{J^2}{q} (\lambda G^{(j,k)}_K(\tau, \tau') + (1-\lambda) G^{(j,k)}_r(\tau, \tau'))^q\right),
\end{align}
where we use bold-face $\mathbf{\Sigma}$ and $\mathbf{G}$ to denote a matrix in the replica indices with entries $\mathbf{\Sigma}_{j,k} = \Sigma^{(j,k)}$. We have also defined the parameter $\lambda \equiv \frac{K}{N}$. The equations of motion are then 
\begin{align}\label{eqn:exactSDeqns}
&\left(\frac{1}{\partial_{\tau} \mathbf{1} - \mathbf{\Sigma_K}}\right)_{j,k} = G_{K}^{(j,k)},\ \ \left(\frac{1}{\partial_{\tau} \mathbf{1} - \mathbf{\Sigma_r}}\right)_{j,k} = G_{r}^{(j,k)},\nonumber \\
& \Sigma^{(j,k)}_I = \Sigma_r^{(j,k)} =  J^2 \left(\lambda G_K^{(j,k)} + (1-\lambda)G_r^{(j,k)}\right)^{q-1},
\end{align}
where to make this expansion well-defined we see that we always need to take the limit $\frac{K}{N} \ll 1/q$ even if we take the limit $q \gg 1$. These equations were also derived in \cite{Zhang:2020us}.

As discussed above, the boundary conditions are such that $G_r$ effectively lives on a larger circle of size $\beta n$ while $G_K$ lives on $n$-copies of a $\beta$-length circle. Explicitly, this is 
\begin{align}\label{eqn:bcs}
    &G_K^{(j,k)}(\beta^-, \tau') = -G_K^{(j,k)}(0^+,\tau'), \nonumber \\
    &G_r^{(j,k)}(\beta^-,\tau') = G_r^{(j+1,k)}(0^+,\tau'),\ 0 \leq j < n-1, \nonumber \\
    & G_r^{(n-1,k)}(\beta^-,\tau') = -G_r^{(0,k)}(0^+,\tau'),
\end{align}
for all $j$ and $k$. There is also the anti-symmetry condition that
\begin{align}
    &G_{r}^{(j,k)}(\tau, \tau') = -G_{r}^{(k,j)}(\tau',\tau), \nonumber \\
    &G_{K}^{(j,k)}(\tau, \tau') = -G_{K}^{(k,j)}(\tau',\tau),
\end{align}
as well as the condition that when the fermions are coincident they square to $1/2$,
\begin{align}
    G_{r}^{(j,j)}(\tau^+, \tau^-) = \frac{1}{2}= G_K^{(j,j)}(\tau^+, \tau^-),
\end{align}
for $0\leq j\leq n-1$. Having set up the problem, we turn now to solving eq. \eqref{eqn:exactSDeqns} in the limit of small $\lambda = K/N$.

\section{Solving the Schwinger-Dyson equations to leading order in $K/N$}\label{sec:SDsolution}
The equations in eq. \eqref{eqn:exactSDeqns} to leading order in $\lambda$ become 
\begin{align}
&\partial_{\tau} \mathbf{G_K}(\tau, \tau') - \left(\mathbf{\Sigma_r}\star \mathbf{G_K}\right)(\tau, \tau') = \mathbf{1} \delta (\tau - \tau'),\nonumber \\
&\partial_{\tau} \mathbf{G_r}(\tau, \tau') - \left(\mathbf{\Sigma_r}\star \mathbf{G_r}\right)(\tau, \tau') = \mathbf{1} \delta (\tau - \tau'), \nonumber \\
&\Sigma^{(j,k)}_K = \Sigma_r^{(j,k)} =  J^2 \left(G_r^{(j,k)}\right)^{q-1} + J^2 \lambda (q-1) \left(G_K^{(j,k)} - G_r^{(j,k)}\right) \left(G_r^{(j,k)}\right)^{q-2}+  \mathcal{O}(\lambda^2).
\end{align}
When we take the star product of two bold-faced variables, we do matrix multiplication in both the continuous time ``indices'' as well as the replica indices. Explicitly, this product is
\begin{align}
    (\mathbf{M} \star \mathbf{N})^{(j,k)}(\tau_1, \tau_2) = \sum_{\ell=0}^{n-1}\int_0^{\beta} d\tau M^{(j,\ell)}(\tau_1, \tau) N^{(\ell, k)}(\tau, \tau_2).
\end{align}
Now we take the ansatz
\begin{align}
&G_r = G_{r,0} + \lambda G_{r,1} + ...\nonumber \\
&G_K = G_{K,0}+\lambda G_{K,1}+...\ .
\end{align}
It is easy to see that to leading order 
\begin{align}
G_{r,0}^{(j,k)}(\tau, \tau') = G_{n\beta}(\tau - \tau'+j\beta - k\beta),
\end{align}
where $G_{n\beta}(\tau - \tau')$ is the thermal two point function at temperature $n\beta$. To find $G_K$ at leading order, we have the equation
\begin{align}\label{eqn:GKeq}
\partial_{\tau} \mathbf{G_{K,0}}- \mathbf{\Sigma_r}^{(0)} \star \mathbf{G_{K,0}} = \mathbf{1} \delta(\tau - \tau'), 
\end{align}
where we need to impose the boundary condition that $G_K^{(j,k)}(\tau, \tau')$ lives on a circle of length $\beta$ for all $j,k$.

\subsection{Solving for $G_{r,1}$}
We can now try to solve for the order $\lambda$ part of $G_r$.  We have
\begin{align}
-\partial_{\tau} \mathbf{1} \star G_{r,1} + \Sigma \star G_{r,1}  + (q-1) J^2 G^{q-2} G_{r,1} \star G  = -(q-1) J^2 (G_K-G) G^{q-2} \star G,
\end{align}
where we have defined $G_{n\beta} \equiv G$. It is convenient to think of $G_{r,1}$ as a function of two times which both live on a single $n\beta$-length circle. In other words,
\begin{align}
G_r^{(j,k)}(\tau, \tau') = G_r(\tau + j\beta, \tau'+ k\beta).
\end{align}
Thus, the star product just means matrix multiplcation in the continuous time indices, but where the intermediate time integrals run from $0$ to $n\beta$. Following the steps in \cite{Maldacena:2016vs}, we can multiply $G$ in from the left on both sides of the equality to get
\begin{align}
(1-K_c) \star G_{r,1} = J^2 (q-1) G \star (G_K-G) G^{q-2} \star G,
\end{align}
where we used the equations of motion on for $G$ and defined the ladder kernel, following \cite{Maldacena:2016vs},
\begin{align}
    K_c(\tau_1, \tau_2; \tau_3, \tau_4) \equiv -J^2 (q-1) G(\tau_{13})G(\tau_{24})G^{q-2}(\tau_{34}).
\end{align}
We remind the reader that the $\star$ operation denotes matrix multiplication where we view $G$, $\Sigma$ as matrices in their arguments $\tau, \tau'$. The time-ordered four point function in SYK, $\frac{1}{N^2} \sum_{i,j=1}^N \braket{\mathcal{T} \psi^i \psi^i \psi^j \psi^j}$, can be written in terms of this ladder kernel \cite{Maldacena:2016vs}. The connected part of the four-point function $\mF$ obeys the equation
\begin{align}\label{eqn:KcF}
   \mF = \frac{1}{1-K_c} \star \mF_0;\ \ \mF_0(\tau_1, \tau_2; \tau_3, \tau_4) = G(\tau_{13})G(\tau_{42}) - G(\tau_{14})G(\tau_{32}).
\end{align}
The connected part of the four point function $\mF(\tau_1, \tau_2;\tau_3,\tau_4)$ inherits its symmetry properties from the four point function itself. Thus, we have
\begin{align}\label{eqn:Fsymm}
&\mF(\tau_1, \tau_2;\tau_3,\tau_4) = -\mF(\tau_2, \tau_1;\tau_3,\tau_4)\nonumber \\
&\mF(\tau_1, \tau_2;\tau_3,\tau_4) = \mF(\tau_3, \tau_4;\tau_1,\tau_2).
\end{align}

Thus, to find $G_{r,1}$ we just need to invert the ladder kernel. Recognizing that the $G$'s on the right are related to $\mF_0$ in eq. \eqref{eqn:KcF}, we land on the formula
\begin{align}\label{eqn:GrpreGI}
G_{r,1}(\tau_1, \tau_2) = \frac{J^2}{2} (q-1) \int d \tau \int d\tau' \mF(\tau_1, \tau_2; \tau, \tau') (G_K(\tau, \tau')-G(\tau, \tau'))G^{q-2}(\tau, \tau').
\end{align}
Note that this formula is exact at large $N$ and so holds for all values of $\beta J$ and $q$. 

\subsubsection*{Solving for $G_K$}
To solve this equation, we need to solve for $G_K$ at order $\lambda^0$. To do that, we write the $G_K$ equation in eq. \eqref{eqn:GKeq} more explicitly as
\begin{align}
\partial_{\tau} G_{K,0}^{(j,k)}(\tau, \tau') - J^2 \sum_{\ell} (G^{(j,\ell)})^{q-1} \star G_{K,0}^{(\ell,k)} = \delta^{j,k} \delta (\tau - \tau'),
\end{align}
where here $\tau, \tau'$ are defined on a single circle of length $\beta$. Since the expressions we wrote above for $G_r$ were defined on a circle of $n\beta$-length, we can introduce a different function for $G_K^{(j,k)}$ which is defined on this circle. Namely, 
\begin{align}\label{eqn:GKnbeta}
G_K(\tau, \tau') \equiv \sum_{j,k} G_K^{(j,k)}(\tau -j, \tau-k) \mathbf{1}_j(\tau) \mathbf{1}_k(\tau'),
\end{align}
with $\mathbf{1}_j(\tau) = 1$ if $j < \tau < j+1$ and zero otherwise.

In terms of this $G_K$, the saddle-point equation is
\begin{align}\label{eqn:nbetaGIeqn}
\partial_{\tau} G_{K,0}(\tau, \tau') - J^2 G^{q-1} \star G_{K,0} = \delta (\tau - \tau') + \sum_j \delta(\tau - j) \disc G_{K,0}(j,\tau'),
\end{align}
where $\disc G_K(j,\tau') = G_K(j^+,\tau') -  G_K(j^-,\tau')$, i.e. it is the discontinuity of $G_K$ in its first argument when transitioning from the $j-1$'th circle to the $j$'th circle. To find a solution, we plug in a simple ansatz for this solution
\begin{align}\label{eqn:GKansatz}
G_K(\tau,\tau') = G(\tau - \tau') + \sum_{\ell m} G(\tau-\ell) c_{\ell m} G(m - \tau'),
\end{align}
where $c_{\ell m}$ is some anti-symmetric matrix which needs to be solved for.\footnote{The ansatz in eq. \eqref{eqn:GKansatz} was motivated by thinking about computing $G_K$ in the replicated theory by the means of inserting twist operators that connect the various replicas. We detail how this works in Appendix \ref{app:SYKtwistop}. } By construction, this ansatz obeys eq. \eqref{eqn:nbetaGIeqn}. The boundary condition that $G_K$ effectively live on $n$ circles each of $\beta$-length,
\begin{align}
G_K(j^+, \tau') = -G_K(j+1^-,\tau'),
\end{align}
then translates to a condition on $c_{\ell m}$:
\begin{align}\label{eqn:clmmain}
\sum_{\ell} (G(j^+,\ell) + G(j+1^-,\ell)) c_{\ell m} = -(\delta_{j,m} + \delta_{j+1,m}).
\end{align}
We will turn to solving this equation for $c_{\ell m}$ momentarily.

We can then take the guess eq. \eqref{eqn:GKansatz} for $G_K$ and plug it back into our eq. \eqref{eqn:GrpreGI} for $G_{r,1}$. We find 
\begin{align}
G_{r,1}(\tau_1, \tau_2) = \frac{J^2}{2} (q-1) \int d \tau \int d\tau' \sum_{\ell m} c_{\ell m} \mF(\tau_1, \tau_2; \tau, \tau') G(\tau, \ell) G(m,\tau') G^{q-2}(\tau, \tau').
\end{align}
We recognize that 
\begin{align}\label{eqn:Kc}
K_c(\ell,m; \tau, \tau') \equiv J^2 (q-1) G(\tau, \ell) G(m,\tau') G^{q-2}(\tau, \tau'), 
\end{align}
and then use that
\begin{align}
K_c \star \mF = K_c \star \frac{1}{1-K_c} \star \mF_0 = -\mF_0(\tau_1, \tau_2; \ell, m) + \mF(\tau_1, \tau_2; \ell, m).
\end{align}
All in all we are left with the simple answer 
\begin{align}\label{eqn:Grgenfinal}
G_{r,1}(\tau_1, \tau_2) = \frac{1}{2}\sum_{\ell m} c_{\ell m} \left(\mF(\tau_1, \tau_2; \ell, m) - \mF_0(\tau_1, \tau_2; \ell,m)\right),
\end{align}
with $c_{lm}$ defined by eq. \eqref{eqn:clmmain}. Eq. \eqref{eqn:Grgenfinal} is the main result of this section and one of the main results of this paper. Note that eqs. \eqref{eqn:clmmain} and \eqref{eqn:Grgenfinal} are exact in $q$ and $\beta \mJ$. We now turn to determining the $c_{\ell m}$ in two regimes where they can be computed analytically.

\subsection{Solving for $c_{\ell m}$ at large $\beta J$}
We now assume that $\beta J$ is large so that the functions $G$ are small away from the coincident limit. This means that in eq. \eqref{eqn:clmmain} for $c_{\ell m}$ we can drop terms where $\ell \neq j$ or $j+1$ since these terms will be proportional to $1/(\beta J)^{2/q}$. Doing so gives us the equation
\begin{align}\label{eqn:clmlargebetaJ}
\frac{1}{2} (c_{j+1m} -c_{jm}) =  \delta_{j+1,m}+\delta_{j,m}.
\end{align}
Using the fact that $c_{jj} = 0$, we see that this is solved by 
\begin{align}
c_{jm} = 2 \sgn(j-m) + \mathcal{O}(1/(\beta J)^{2/q}).
\end{align}
Thus for large $(\beta J)^{2/q}$, $G_{r,1}$ is given by 
\begin{align}\label{eqn:GrlargebetaJ}
G_{r,1}(\tau_1, \tau_2) = 2\sum_{\ell > m}^{n-1} \left( \mF(\tau_1, \tau_2;\ell,m) - \mF_0(\tau_1, \tau_2; \ell, m)\right).
\end{align}

\subsection{Solving for $c_{\ell m}$ at large $q$}
At large $q$, the theory is free (to leading order in $q$), and so the expression for $c_{\ell m}$ given in eq. \eqref{eqn:clmmain} just becomes
\begin{align}
\frac{1}{2} \sum_{\ell} \left(\sgn(j^+-\ell) + \sgn(j+1^- -\ell)\right) c_{\ell m} = -(\delta_{j,m}+ \delta_{j+1,m}).
\end{align}
Note that here we are assuming that $\frac{1}{q} \log (\beta \mJ)$ is small.
One can check that this can be solved by 
\begin{align}\label{eqn:clmlargeq}
c_{\ell m} = \frac{1}{2} \left( \delta_{\ell, m+1} - \delta _{m,\ell+1}\right) + \mathcal{O}(1/q).
\end{align}
Thus for large $q$, the solution is given by 
\begin{align}\label{eqn:Gr1largeq}
G_{r,1}(\tau_1, \tau_2) = \frac{1}{2}\sum^{n-1}_{\ell=0} \left( \mF(\tau_1, \tau_2; \ell + 1, \ell) - \mF_0(\tau_1 , \tau_2; \ell+1,\ell)\right).
\end{align}
In principle, one could correct the solutions in eqs. \eqref{eqn:GrlargebetaJ} and \eqref{eqn:Gr1largeq} by solving eq. \eqref{eqn:clmmain} perturbatively in $1/q$ or $1/(\beta J)$.

\section{Analytic continuation}\label{sec:analyticcontinuation}
As discussed in Sec. \ref{sec:setup}, the goal is to compute the modular-flowed correlation function, which will tell us directly about the position of the quantum extremal surface. This can be done by continuing 
\begin{align}
    \frac{1}{N-K} \sum_{i=1}^{N-K} \braket{\psi_L^i(\tau) \left(\rho_{LI}^{k} \otimes \rho_r^{n-1-k}\right) \psi_R^i(\tau')}_{\beta}
\end{align}
away from integer $n$ and $k \leq n$ and then taking $n \to 1$ and $k\to is$. Furthermore, we continue $\tau \to iT_L$ and $\tau' \to -iT_R$ to get the correlator in eq. \eqref{eqn:intnmodflowcorr}. This requires us to put one of the fermions on the $k$'th replica and the other on the first replica. This means we want to compute 
\begin{align}
    G_r(\tau + k, \tau') =  \frac{1}{N-K} \sum_{i=1}^{N-K} \frac{\braket{\psi_L^i(\tau) \left(\rho_{LK}^{k} \otimes \rho_r^{n-1-k}\right) \psi_R^i(\tau')}_{\beta}}{\text{Tr} \rho_{LK}^{n}}
\end{align}
to leading order in $\lambda = K/N$ and at replica number $n$. Here $1/2> \tau'>0$ and $1> \tau >1/2$ since $\psi_L$ is an operaor on the left. Again we have set $\beta =1$. As discussed in Sec. \ref{sec:setup}, we are only interested in the order $\lambda$ piece that is also exponentially growing in $-T_R$ as we move the fermion insertion back in time. Such exponentially growing terms only arise from out-of-time order or ``crossed" configurations between the four times in the function $\mF$ discussed above. We turn now to continuing these sums in two limits where $\mF$ is known: the large $\beta J$ and large $q$ limits.

\subsection{Analytic continuation at large $\beta J$}
Here we continue the answer for $G_r$ at large $\beta J$ given in eq. \eqref{eqn:GrlargebetaJ}. As we found in the previous section
\begin{align}\label{eqn:GrlargebetaJ2}
    G_r(\tau, \tau') = G_{n}(\tau - \tau') + \frac{2K}{N} \sum_{\ell > m} \left(\mF(\tau, \tau'; \ell,m) - \mF_0(\tau, \tau';\ell,m)\right) + \mathcal{O}((K/N)^2).
\end{align}
Setting $\tau \to \tau + k$ and then continuing $\tau, \tau' \to 1/2 + iT_L,\,-iT_R$ respectively, we see that the first term is trivial to continue in $n$ and $k$ since it is just given by the usual thermal two point function. Taking $n\to 1$ and $k \to is$, we have
\begin{align}\label{eqn:Gn=1}
    G_n(\tau +k+1/2-\tau') \to G_1(\tau +1/2 +is - \tau') = b \left(\frac{\pi}{\cosh(T_L - T_R +s +i\delta)}\right)^{2\Delta},
\end{align}
where $b$ is the overall normalization in the two point function that is defined by the equation \cite{Maldacena:2016vs}.
\begin{align}\label{eqn:bdef}
    b^q \pi = \frac{\left(1/2 - \Delta \right) \tan \pi \Delta}{(\beta J)^2},\ \Delta = \frac{1}{q}.
\end{align}
Note that $b \sim 1/(\beta J)^{2/q}$ in the large $\beta J$ limit. We will always work to leading non-trivial order in $b$.

Analytically continuing the second term on the right hand side of eq. \eqref{eqn:GrlargebetaJ2} is less trivial, although similar calculations were done in \cite{Chandrasekaran:2021wx}. We outline the general method of calculation. Since we are only interested in the crossed configurations which grow exponentially in $T_R$, we can restrict the sum over $\ell$ and $m$ and also drop the term involving a sum over $\mF_0$ since this will not grow exponentially in $T_R$ for any orderings. Dropping the non-crossed orderings gives
\begin{align}\label{eqn:crossedterms}
    \left. \sum_{\ell > m} \mF(\tau+k, \tau';\ell,m)\right \vert_{\text{crossed}} = \sum_{\ell = 1}^{k} \mF(\tau + k,\tau';\ell, 0) + \sum_{\ell= k+1}^{n-1}\sum_{m=1}^{k} \mF(\tau + k, \tau';\ell, m).
\end{align}
We now analytically continue the single and double sums on the right separately using the standard methods of \cite{Faulkner:2014aa, Balakrishnan:2019ti}.
\subsubsection*{Single sum}
To continue the single sum, we write the sum as a contour integral as
\begin{align}\label{eqn:sumcontour}
    \sum_{\ell =1}^k \mF(\tau+k, \tau';\ell,0) = \frac{1}{i}\oint_{\mathcal{C}} ds' \left(\frac{1}{e^{2\pi s'}-1} - \frac{1}{e^{2\pi s'/n}-1}\right) \mF(\tau +k,\tau';-is',0)
\end{align}
where the contour $\mathcal{C}$ encircles the poles of the kernel at $s' = i\ell$ for $\ell = 1,...,k$. Now as a function of $s'$, $\mF(\tau + k, \tau';-is',0)$ has branch cuts at $-is' = \tau +k$ and $-is = \tau'$. We can unwrap the contour and write it as two real contours, being careful not to cross the branch cuts. For an illustration of this idea see Figure 1 of \cite{Faulkner:2014aa}. If we set $\tau = 1/2 + iT_L$ and $\tau = \delta - iT_R$ where $1/(J) \ll \delta \ll 1/2$, then we can unwrap the contour so that $\mathcal{C} \to \mathcal{C}_+ \cup \mathcal{C}_-$ where the upper contour $\mathcal{C}_+$ sits at $\text{Im} s' = 1/2^- + k$ and the lower contour sits at $\text{Im}s' =1/2$. Note that we have not passed any branch cuts.

On the lower contour, $\mC_-$, the kernel integrated against the four point function in eq. \eqref{eqn:sumcontour} takes the form
\begin{align}
    \left(\frac{-1}{e^{2\pi s'}+1} - \frac{1}{e^{2\pi (s'+i/2)/n}-1}\right)
\end{align}
which, upon analytic continuation to non-integer $n$, vanishes at $n \to 1$ and for all $k$. Thus, we can ignore this term in the $n\to 1$ limit.

The contribution from the upper contour $\mC_+$ gives
\begin{align}
    \frac{-1}{i} \int ds \left(\frac{-1}{e^{2\pi (s'+ik)}+1}- \frac{1}{e^{2\pi (s'+i/2 +ik)/n}-1}\right) \mF(\tau+k,\tau'; -is'+k +1/2,0).
\end{align}
Now we need to continue this expression in $k$ and $n$. The correct answer for this was found in \cite{Faulkner:2014aa,Balakrishnan:2016ttg} and the prescription is to set $e^{2\pi ik} =1$ while setting $e^{2\pi ik/n} \to e^{-2\pi s}$ if we take $k \to is$ and $n\to 1$. The final answer can then be written
\begin{align}\label{eqn:finalsinglesumunint}
    \sum_{\ell =1}^k \mF(\tau+k, \tau';\ell,0) \to \frac{i\pi}{2}\int_0^s dt \int_{-\infty}^{\infty}ds' \frac{\mF(\tau+is,\tau';-is'+is+1/2,0)}{\cosh^2(\pi (s'-t))},
\end{align}
for $k \to is$ and $n\to 1$. 

At large $\beta J$ the form of the exponentially growing part of $\mF$ is known \cite{Qi:2018wg, Maldacena:2016vs}. In the time-ordering that is relevent to us, namely $\tau_1>\tau_3>\tau_2>\tau_4$, the answer is
\begin{align}\label{eqn:Fexpression}
\mF(\tau_1, \tau_2; \tau_3, \tau_4) = -\frac{\beta \mJ F_d}{2 \alpha_S q^2 \pi } \frac{\sin \left(\pi( \tau_{12}^+ -\tau_{34}^+)\right)}{\sin \left ( \pi \tau_{12}^- \right) \sin \left( \pi \tau_{34}^- \right)},
\end{align}
where $\alpha_S$ is the coupling of the Schwarzian mode, defined in \cite{Maldacena:2016vs} and $F_d = G(\tau_{12}) G(\tau_{34})$.
Setting $\tau = 1/2 + iT_L$ and $\tau' = -iT_R$, the numerator of the integrand in eq. \eqref{eqn:finalsinglesumunint} becomes
\begin{align}
    &\mF(\tau+is,\tau';-is'+is+1/2,0) = -i\frac{b^2 \pi^{4\Delta} \beta \mJ}{2\alpha_sq^2 \pi} \frac{\sinh(\pi(T_L-T_R +s'))}{\cosh^{1+2\Delta}(\pi (T_L+T_R+s)) \cosh^{1+2\Delta}(\pi(s'-s))}\nonumber \\
    & \approx -i\frac{b^2 \pi^{4\Delta} \beta \mJ}{4\alpha_sq^2 \pi} \frac{e^{\pi(T_L - T_R+s')}}{\cosh^{1+2\Delta}(\pi (T_L+T_R+s)) \cosh^{1+2\Delta}(\pi(s'-s))}
\end{align}
where in the second line we took $T_L - T_R$ large.

To evaluate eq. \eqref{eqn:finalsinglesumunint}, we thus need to perform the integral
\begin{align}
\int_0^s dt \int_{-\infty}^{\infty}ds' \frac{e^{\pi s'}}{\cosh^2(\pi (s'-t))\cosh^{1+2\Delta}(\pi(s'-s))} =\frac{1}{\pi} \int_{-\infty}^{\infty}ds' \frac{f_1(s',s)}{\cosh^{1+2\Delta}(\pi s')}
\end{align}
with
\begin{align}
    f_1(s',s) = e^{\pi(s'+s)}f_0(s',s),\ \ f_0(s',s) = \tanh(\pi(s+s'))-\tanh(\pi s').
\end{align}
All in all, we have that the single sum continues to
\begin{align}\label{eqn:finalsinglesum}
    &\sum_{\ell =1}^k \mF(\tau+k, \tau';\ell,0) \to
    C(\beta \mJ, q)\frac{ e^{\pi(T_L - T_R)}}{\cosh^{1+2\Delta}(\pi (T_L+T_R+s))} \int_{-\infty}^{\infty} ds' \frac{f_1(s',s)}{\cosh^{1+2\Delta}(\pi s')},
\end{align}
where we have defined the constant
\begin{align}\label{eqn:Cdef}
    C(\beta \mJ, q) = \frac{b^2 \pi^{4\Delta} \beta \mJ}{8\alpha_sq^2 \pi}.
\end{align}
We will leave evaluating this final integral until after combining with the double sum.

\subsubsection*{Double sum}
We now continue the double sum in eq. \eqref{eqn:crossedterms}. Our method is the same as for the single sum but now we introduce two contour integrals, unwrap the contours (being careful not to cross branch-cuts), set $e^{2\pi ik}=1$ and drop terms that vanish at $n=1$. Sparing the details, we find
\begin{align}
    &\sum_{\ell= k+1}^{n-1}\sum_{m=1}^{k} \mF(\tau + k, \tau';\ell, m) \to \nonumber \\
    & \frac{\pi^2}{4} \int_0^s dt_1 dt_2 \int_{-\infty}^{\infty}ds_1 ds_2 \times \frac{\mF(\tau+is, \tau';-is_1 + is +1/2^+, -is_2+is+1/2^-)}{\cosh^2(\pi (s_1-t_1))\cosh^2(\pi (s_2-t_2))}
\end{align}
as $k \to is$ and $n\to 1$.

The connected part of the four point function takes the form 
\begin{align}
    &\mF(\tau+is,\tau';-is_1+is+1/2^+,-is_2+is+1/2^-) \nonumber\\
    &=-4C(\beta \mJ , q) \frac{\sin(\pi(i(T_R-T_L-s_1-s_2+s)+\frac{\pi}{2} )}{\sin^{1+2\Delta}(i\pi (T_L+T_R+s)+\pi/2) \sin^{1+2\Delta}(\pi(is_2-is_1+\epsilon))}\nonumber \\
    &  \to -2 C(\beta \mJ, q) \frac{e^{\pi(T_L-T_R+s_1+s_2-s)}}{\cosh^{1+2\Delta}(\pi (T_L+T_R+s)) \sinh^{1+2\Delta}(\pi(is_2-is_1+\epsilon))},
\end{align}
where again in the last line we took the large $T_L - T_R$ limit. In this limit, we then need to compute the integral
\begin{align}
    & \int_0^s dt_1 dt_2 \int_{-\infty}^{\infty}ds_1 ds_2 \times \frac{e^{\pi(s_1+s_2-s)}}{\cosh^2(\pi (s_1-t_1))\cosh^2(\pi (s_2-t_2))\sinh^{1+2\Delta}(\pi(is_2-is_1+\epsilon))}.
\end{align}
With a view toward combining this double sum term with the single sum term, we shift $s_2 \to s_2-i/2+i\epsilon+s_1$ so that we find this integral to be
\begin{align}
    & \frac{1}{\pi^2}\int_{-\infty}^{\infty}ds_2 \frac{f_2(s_2,s)}{\cosh^{1+2\Delta}(\pi s_2)}.
\end{align}
where we have defined
\begin{align}
   f_2(s_2-i/2,s)= \pi^2 \int_0^s dt_1 dt_2 \int_{-\infty}^{\infty} ds_1 \frac{e^{\pi(-2s_1-s_2-s)}}{\cosh^2(\pi (s_1+t_1))\cosh^2(\pi (s_2+s_1+t_2+i\epsilon))}.
\end{align}
To compute this, we do the two $t$-integrals first to land on
\begin{align}
   &f_2(s_2-i/2,s)= \int_{-\infty}^{\infty} ds_1\  e^{\pi(-2s_1-s_2-s)} f_0(s_1,s)f_0(s_1+s_2,s).
\end{align}
This integral can be done in Mathematica or by contour integration.

Thus, all in all we get that the double sum becomes
\begin{align}
      &\sum_{\ell= k+1}^{n-1}\sum_{m=1}^{k} \mF(\tau + k, \tau';\ell, m) \to \nonumber \\
     & -C(\beta \mJ,q) \frac{e^{\pi(T_L - T_R)}}{\cosh^{1+2\Delta}(\pi (T_L+T_R+s))} \int_{-\infty}^{\infty} \frac{\frac{1}{2}f_2(s_2,s)}{\cosh^{1+2\Delta}(\pi s_2)}
\end{align}

\subsubsection*{Combining the single and double sums}
To get the final answer, we must add the two contributions we have just computed. Doing so gives us the final answer
\begin{align}\label{eqn:modflowcorrunint}
  \frac{1}{N-K} \sum_{i=1}^{N-K} \braket{\psi_L(T_L) \Delta_{LI}^{is} \psi_R(T_R)} = &b \left(\frac{\pi}{\cosh(T_L + T_R +s)}\right)^{2\Delta} \nonumber \\
  &+ \frac{2K}{N} \frac{ C(\beta \mJ, q)e^{\pi (T_L + T_R)}}{\cosh^{1+2\Delta}(\pi (T_L + T_R+s))} \int_{-\infty}^{\infty}ds' \frac{f_1(s',s)- \frac{1}{2} f_2(s',s)}{\cosh^{1+2\Delta}(\pi s')}
\end{align}
with $C(\beta \mJ, q)$ defined in eq. \eqref{eqn:Cdef}.

We find that the combination simplifies considerably to 
\begin{align}
    f_1(s',s) - \frac{1}{2} f_2(s',s) = \frac{\sinh(\pi s)}{\cosh(\pi s')},
\end{align}
where we have dropped terms which are anti-symmetric in $s'$ since this kernel will be integrated against an even function in $s'$ when inserted into eq. \eqref{eqn:modflowcorrunint}. Now, using the fact that
\begin{align}
    \int_{-\infty}^{\infty} ds' \frac{1}{\cosh^{2+2\Delta}(\pi s')} = \frac{\Gamma(\Delta +1)}{\sqrt{\pi} \Gamma(\Delta + 3/2)},
\end{align}
we get the remarkably simple answer that 
\begin{align}
  \frac{1}{N-K} \sum_{i=1}^{N-K} \braket{\psi_L(T_L) \Delta_{LI}^{is}\Delta_L^{-is} \psi_R(T_R)} = &b \left(\frac{\pi}{\cosh(\pi(T_L + T_R))}\right)^{2\Delta} \nonumber \\
  &+ \frac{K}{N} \frac{b\pi^{2\Delta}\Delta C_{\Delta}(\beta \mJ, q)e^{\pi (T_L - T_R)}}{\cosh^{1+2\Delta}(\pi (T_L + T_R))} \left(e^{2\pi s}-1\right)
\end{align}
where 
\begin{align}
C_{\Delta}(\beta \mJ, q) = \frac{ \Gamma(\Delta)}{b\pi^{2\Delta+1/2} \Gamma(3/2+\Delta)}C(\beta \mJ,q).
\end{align}

We can also suggestively write this as
\begin{align}\label{eqn:allorderslambda}
   & \frac{1}{N-K} \sum_{i=1}^{N-K} \braket{\psi_L(T_L) \Delta_{LI}^{is}\Delta_L^{-is} \psi_R(T_R)} \nonumber \\
    &= b \left(\frac{\pi}{\cosh(\pi(T_L +T_R)) - \frac{K}{2N}C_{\Delta} e^{\pi(T_L - T_R)} (e^{2\pi s}-1) }\right)^{2\Delta}
\end{align}
expanded out to leading order in $K/N\times C_{\Delta}$.

As discussed in Sec. \ref{sec:intro}, the fact that the correction to this correlator has an $s$ dependence which is entire on the complex $z = e^{-2\pi s}$ plane is a smoking gun signal of the appearance of a quantum extremal surface. In the words of \cite{Boer:2019td}, this means that for SYK at large $\beta \mJ$, the modular chaos bound is saturated and maximal modular chaos is intimately tied up with the physics of QES's. Shiftin $T_R \to T_R+s$, we see agreement between eq. \eqref{eqn:allorderslambda} and eq. \eqref{eqn:maxmodchaos} in the introduction. By comparing with eq. \eqref{eqn:maxmodchaos} we can read off the exact position of the putative QES. We find the prediction 
\begin{align}\label{eqn:mainqesresult}
    \delta x_Q^+ = \frac{K}{N}C_{\Delta}(\beta \mJ, q) =  K \frac{ \beta \Delta}{2\pi \phi_r} \frac{b \pi^{2\Delta+1/2}\Gamma(\Delta+1)}{2\Gamma(\Delta + 3/2)},
\end{align}
where we have identified $N\alpha_s/\mJ = \frac{\phi_r}{2\pi}$ with a putative boundary value of a dilaton, $\phi_r$, in the dual theory. We have also used $\Delta = 1/q$ and remind the reader that $b \sim 1/(\beta J)^{2/q}$ in the large $\beta J$ limit.

\subsection{Analytic continuation at large $q$}
Another limit where we can compute some quantities analytically is the large $q$ limit. As computed in eq. \eqref{eqn:Gr1largeq}, we found that
\begin{align}\label{eqn:Grlargequncont}
    G_r(\tau, \tau') = G_{n}(\tau - \tau') + \frac{K}{2N} \sum_{j=0}^{n-1} \left(\mF(\tau, \tau'; j+1,j) - \mF_0(\tau, \tau';j+1,j)\right) + \mathcal{O}((K/N)^2).
\end{align}
As in the large $\beta \mJ$ limit we would like to continue $G_r(\tau + k,\tau')$ in $n$ and $k \leq n$. Note that when $n=1$, this sum vanishes because $\mF(\tau, \tau'; 1,0)\vert_{n=1} = 0$. Unfortunately, because we want to continue in both $n$ and $k$ simultaneously, we found it hard to find a correct continuation of this sum which vanishes as $n\to 1$ and agrees with eq. \eqref{eqn:Grlargequncont} for integer $n>1$. We were, however, able to find a continuation in $n$ of the quantity $G_r(\tau + n-1, \tau')$. This quantity tells us about the so-called modular energy since 
\begin{align}
    G_r(\tau + n-1,\tau') =\frac{ \braket{\psi(\tau) \rho_{LI}^{n-1} \psi(\tau')}_{\beta}}{\braket{\rho_{LI}^{n-1}}_{\beta}}.
\end{align}
We see that continuing in $n$ and stripping off the order $n-1$ piece of $G_r$ gives us information about modular energy $H_{LI} = -\log \rho_{LI} $.

As in the previous sub-section, we will only be interested in the parts of $G_{r,1}$ which grow exponentially in $T_R$ once we continue $\tau, \tau'$ off the real axis. This means that we can ignore the non-crossed configurations which contribute to $G_{r,1}(\tau+n-1, \tau')$. Dropping such contributions, as well as the contributions to $G_{r,1}$ from $\mF_0$ we see that we have 
\begin{align}\label{eqn:crossedconfigs}
G_{r,1}(\tau + n-1, \tau')\vert_{\text{crossed}} &= \mF(\tau + n-1, \tau'; n, n-1) + \mF(\tau + n-1, \tau'; 1,0)\nonumber \\
&= \mF(\tau + n-1, \tau'; n-1,0) + \mF(\tau + n-1, \tau'; 1,0),\ \ n \geq 2,
\end{align}
where in the second line we used the symmetry properties of $\mF$ in eq. \eqref{eqn:Fsymm} to re-write the first term. Note, however, that such crossed terms \emph{do not contribute} to $G_{r,1}(\tau, \tau')$ when $n=1$. Thus, the expression in eq. \eqref{eqn:crossedconfigs} is only valid for $n \geq 2$. We thus need to continue this expression in such a way that it is equal to this for $n\geq 2$ but is zero when $n = 1$. Unfortunately, the \enquote{naive} continuation of the two terms in eq. \eqref{eqn:crossedconfigs} to $n = 1$ (just take $n$ and make it complex) does not vanish at $n =1$.

With this goal in mind, and after much trial and error, we consider re-writing these two terms as 
\begin{align}\label{eqn:threeterms}
2\sum_{j=1}^{n-1} \mF(\tau + n-1, \tau'; j, 0) - \sum_{j=1}^{n-2} \mF(\tau + n-1, \tau'; j, 0) -\sum_{j=2}^{n-1} \mF(\tau + n-1; j,0).
\end{align}
While this looks at first glance like a more complicated expression, such sums can be continued in $n$ in a way analogous to the previous sub-section. First, we write the latter two terms as 
\begin{align}
&\sum_{j=1}^{n-2} \mF(\tau + n-1, \tau'; j, 0) + \sum_{j=2}^{n-1} \mF(\tau + n-1; j,0) \nonumber \\
& =\sum_{j=1}^{n-2} \mF(\tau + n-1, \tau'; n-1-j, 0) + \sum_{j=1}^{n-2} \mF(\tau + n-1; j+1,0).
\end{align}
As before, we write these sums as contour integrals, where we integrate against the kernel
\begin{align}
k_n(s) = \frac{1}{i} \left( \frac{1}{e^{2\pi s}-1} - \frac{1}{e^{2\pi s/n}-1}\right).
\end{align}
Note that $k_n(s)$ has poles at $s = ik$ for $k =1,...,n-1$ with residue one and has a pole of residue $n-1$ at $s = 0$. Thus, when we unwrap the contour integrals, we want to deform the contours such that as we continue in $n$ we never cross a branch cut. In order to get to a configuration like that we have to deform the lower rail down past the $s=0$ pole, picking it up. Thus, we have
\begin{align}
&\sum_{j=1}^{n-2} \mF(\tau + n-1, \tau'; j+1, 0) = (n-1) \mF(1,0) \nonumber \\
&+ \int ds' k_n(s'-i/2) \mF( -is' + 1/2,0) - \int ds' k_n(s'-3i/2) \mF(-is' + n-1/2,0)
\end{align}
where for simplicity we are going to assume that $1 > \tau \geq 1/2$. Furthermore, we have introduced new notation where $\mF(\sigma, \sigma') \equiv \mF(\tau +n-1, \tau';\sigma, \sigma')$, where we have dropped the explicit dependence on $\tau, \tau',\  n$ since these will remain the same throughout the calculation.

Similarly, 
\begin{align}
&\sum_{j=1}^{n-2} \mF(\tau + n-1, \tau'; n-1-j, 0) = (n-1) \mF(n-1,0) \nonumber \\
&+ \int ds' k_n(s'-i/2) \mF(n-1/2 + is' ,0) - \int ds' k_n(s'-3i/2) \mF(is' +1/2,0).
\end{align}
Now, since $k_n(s) \to 0$ as $n\to 1$, every term in this expression goes to zero as $n \to 1$. Strictly speaking, in order to continue the terms like $\mF(\tau + n-1, \tau'; n-1,0)$ in $n$, one would naively have to cross a branch cut as the third operator would cross the second operator when $n -1 = \tau'$. What we will do instead is to just declare the continuation to be the $n \geq 2$ expression continued down to $n=1$. This amounts to moving off the principle sheet of the function $\mF$.

Plugging these expressions into eq. \eqref{eqn:threeterms} and continuing the first sum in eq. \eqref{eqn:threeterms} in the standard way, we see that as $n \to 1$ all the integral terms cancel off each other and we are just left with the two pole terms. We then have
\begin{align}
&\lim_{n \to 1} \frac{1}{n-1}G_{r,1}(\tau+n-1, \tau') \nonumber\\
& = \frac{-1}{2} \left( \mF(\tau + n-1, \tau'; 1,0) + \mF(\tau + n-1, \tau'; n-1,0)\right)\vert_{\text{continued}} + \mathcal{O}((n-1)^2),
\end{align}
where the sub-script ``continued'' signifies that we have continued $\mF$ off its principal sheet.

We would like to explicitly evaluate this expression. The formula for $\mF$ at large $q$ is actually known for all $\beta \mJ$. In the order $\tau_1 > \tau_3>\tau_2>\tau_4$, it was computed in \cite{Streicher:2019vn} to be
\begin{align}\label{eqn:largeqF}
    \frac{\mF(\tau_1,\tau_2;\tau_3,\tau_4)}{G(\tau_{12}) G(\tau_{34})} = - \frac{2 \sin \left(\frac{\phi_{12}^+ - \phi_{34}^+}{2}\right)}{\cosh \frac{\pi v}{2} \sin \frac{\phi_{12}^-}{2} \sin \frac{\phi_{34}^-}{2}},
\end{align}
where
\begin{align}\label{eqn:phidef}
    \phi^{\pm}_{ij} = (1-v) \pi + v \frac{2\pi \tau^{\pm}_{ij}}{\beta},\ \  \beta \mJ = \frac{\pi v}{\cos \frac{\pi v}{2}}.
\end{align}
We are using the notation that $\tau_{ij}^{\pm} = \tau_i \pm \tau_j$, and we have dropped terms in the expression for $\mF$ that do not grow exponentially with $T_L-T_R$ upon continuation $\tau \to \tau+iT_L$ and $\tau' \to \tau'-iT_R$.

Setting $\tau \to 1/2 + iT_L$ and $\tau' \to -iT_R+\tau'$, we see that for $n\geq 2$, we have
\begin{align}
&\frac{1}{n-1}G_{r,1}(\tau+n-1, \tau') \nonumber\\
& = \frac{1}{4\cos\left( \frac{\pi v}{2}\right)\sin \left(\frac{\pi (1-v)}{2n} + \pi \frac{v}{n}(n-1/2 + i(T_L+T_R))\right)}\times\nonumber \\
& \left(\frac{\sin\left(\frac{\pi (1-v)}{2n} + \frac{\pi v}{n}(n-3/2 + i (T_L-T_R))\right)}{\sin\left(\frac{\pi (1+v)}{2n} \right)}+ \frac{\sin\left(\frac{\pi (1-v)}{2n} + \frac{\pi v}{n}(1/2+ i (T_L-T_R))\right)}{\sin\left(\frac{\pi (1-v)}{2n} + \frac{\pi v}{n}(n-1) \right)}\right).
\end{align}
As $n\to 1$ this becomes
\begin{align}
&\lim_{n\to1}\frac{1}{n-1}G_{r,1}(\tau+n-1, \tau')
= \frac{ e^{i\pi v/2} e^{\pi v(T_L - T_R)}}{4\cos\left(\frac{\pi v}{2}\right)\cosh(\pi v (T_L +T_R))} + \mathcal{O}(e^{-\pi v(T_L-T_R)})
\end{align}
where we have also taken the large $T_L - T_R$ limit.

All in all, we end up with an answer for the modular energy expectation value of the form: 
\begin{align}\label{eqn:modenergylargeq}
&\frac{1}{N-K}\sum_{i=1}^{N-K} \braket{\psi_R(T_R-i\delta)\log \rho_{LI} \psi_R(T_R+i\delta)}_{\beta} \nonumber \\
&= \frac{-2\pi v}{q\sin(2\pi v \delta +  \frac{\pi (1- v)}{2})} \left(1-\frac{qK}{8\pi N v\cos(\frac{\pi v}{2})}e^{-2\pi v T_R} + (\text{sub-leading}) \right) \nonumber \\
& \to \frac{-2\pi}{q\sin(2\pi \delta)} \left(1-\frac{qK \beta \mJ}{8\pi^2 N}e^{-2\pi T_R} + (\text{sub-leading in }e^{2\pi T_R}) \right)
\end{align}
where in the last line we took the large $\beta \mJ$ or $v \to 1$ limit and used eq. \eqref{eqn:phidef}. Given the discussion in Sec. \ref{sec:intro}, we find a prediction for the QES position at large $q$ and large $\beta \mJ$:
\begin{align}
    \delta x_Q^+ = \frac{qK \beta \mJ}{8\pi^2 N} =  \frac{K\Delta \beta}{16\pi \phi_r},
\end{align}
where we used again that $N\alpha_S/\mJ = \frac{\phi_r}{2\pi}$, with $\phi_r$ the boundary value of the dilaton in some putative holographic dual. We turn now to matching the prediction for a QES position in eq. \eqref{eqn:mainqesresult} with a simple bulk model.

\section{Finding the quantum extremal surface in the bulk}\label{sec:bulkpicture}
In this section, we attempt to match the previous answer to a bulk picture where there is a QES outside the horizon at some location. We will guess a bulk model and then see if it predicts the correct QES location.

The model we propose is that of JT gravity coupled to $N$ free, massive Majorana fermions. We will assume that their mass is set by the boundary conformal dimension according to the usual rules of AdS/CFT. Following the conventions of \cite{AMM}, we take the bulk action to be
\begin{align}
    S = \frac{1}{4\pi} \int d^2 x \sqrt{-g} \left( \phi R + 2 (\phi - \phi_0)\right) + S_{\text{fermions}}
\end{align}
where $S_{\text{fermions}}$ is given in Appendix \ref{app:bulkfermion}.

To find the QES, we need to be able to do a bulk entropy calculation for $N$ free, massive fermions propagating in an AdS$_2$ background. Now, as a guess of the bulk picture, we imagine that the $K$ fermions on the right are dual to a very small entanglement wedge, very close to the right boundary. In the large $\beta \mJ$ limit, one might guess that this boundary is only a thermal time in length. This is because after one thermal time (i.e. one commutator with the Hamiltonian) a fermion in the $K$-subset moves mostly out of $K$, since we are working in the limit $K \ll N$.  

The natural guess is to then compute the entanglement entropy of an interval in AdS$_2$ where one of the end-points is very near the boundary. The complementary region is then the union of a large interval $\boldsymbol{\ell}$, which ends on the left boundary, and a small interval $\boldsymbol{i}$ which ends on the right boundary. This was illustrated in Fig. \ref{fig:bulkpicture}. We would thus like to compute the path integral with two twist fields, one near the boundary and one somewhere else, perhaps at the bifurcation surface. This is isometric to a situation where the twist operators are both off near the boundary. 

Our guess is to look for a QES close to the bifurcation surface. To find such a QES, we need to extremize the generalized entropy for the dilaton solution
\begin{align}
    ds^2 = \frac{-4dx^+ dx^-}{(x^+ x^-+1)^2}, \ \  \phi = \phi_0 + \frac{2\pi \phi_r}{\beta} \frac{1-x^+ x^-}{1+x^+ x^-},
\end{align}
where $x^{\pm}$ are Kruskal coordinates. Letting $S_f(\boldsymbol{\ell i})$ denote the bulk entanglement entropy of a Majorana fermion for the region $\boldsymbol{\ell i}$. Thus, we want to extremize
\begin{align}
    S_{gen}(x^+,x^-) = \phi(x^+,x^-) + K S_f(\boldsymbol{\ell i}) + (N-K) S_f(\boldsymbol{\ell})
\end{align}
where we are \emph{only} including the entanglement between $\boldsymbol{\ell}$ and $\boldsymbol{i}$ due to the $K$ flavors. The remaining $N-K$ flavors just contribute to the overall entropy.

We will look for a solution on the $t=0$ slice at spatial coordinate $x$ and near the bifurcation surface at $x^+ =x^- = 0$. Extremizing the generalized entropy and using the fact that $\partial_{x} S_f(\boldsymbol{\ell}) = \partial_x^2 S_f(\boldsymbol{\ell}) = 0$ by the SL(2) symmetries of AdS$_2$, we get
\begin{align}\label{eqn:dxS'}
    S'_{gen}(x)= 0 \implies \delta x^+_Q = -K \frac{S_f'(\boldsymbol{\ell i})}{\phi''(0)} = -\frac{K \beta }{8 \pi \phi_r}S_f'(\boldsymbol{\ell i}),
\end{align}
where the primes denote derivatives with respect to $x = \frac{x^+ - x^-}{2}$. Note that by symmetry of the thermofield double as well as strong sub-additivity (SSA), $S'(\boldsymbol{\ell i})= I'(\boldsymbol{\ell}:\boldsymbol{i}) \leq 0$ which implies $\delta x_Q^+ \geq 0$. This must be the case if the entanglement wedge is going to be outside the horizon, as it needs to be by entanglement wedge nesting. The fact that SSA implies entanglement wedge nesting is no surprise, see \cite{Akers:2016aa,Koeller:2015qmn,Balakrishnan:2017aa}.

Now we just need to compute $S'_f(\boldsymbol{\ell i})$. In Appendix \ref{app:dS}, we show how to compute $S_f(\boldsymbol{\ell i})$ perturbatively in the size of the region associated to $\boldsymbol{i}$. Indeed, if $\boldsymbol{i}$ is an interval near the right boundary, with endpoint at $t=0$ and Poincare coordinate $z_{\boldsymbol{i}}$. Then the answer we find is
\begin{align}
    S(\boldsymbol{\ell i}) = S(\boldsymbol{\ell}) + S(\boldsymbol{i}) + \delta S(\boldsymbol{\ell i})
\end{align}
where
\begin{align}\label{eqn:dSapprox}
    \delta S(\boldsymbol{\ell i}) \sim -\left(z_{\boldsymbol{i}}\right)^{2\Delta} + \mathcal{O}(z_{\boldsymbol{i}}^{4\Delta}),
\end{align}
with $\Delta$ related to the mass of the bulk fermions and where the constant of proportionality is positive and depends only on $\Delta$.\footnote{Note that by sub-additivity of the entropy, $\delta S(\boldsymbol{\ell i})$ is negative.} We have dropped terms here of order $z_{\boldsymbol{i}}^{4\Delta}$ and higher. This formula follows from a standard procedure of computing the entropy for $\boldsymbol{\ell i}$ via twist operator insertions, as detailed in \cite{Agon:2015ftl}. Since the twist operators are far from each other, we are in an OPE like limit and so can expand in the lightest operators. Eq. \eqref{eqn:dSapprox} represents the contribution from the lightest operator.  In Appendix \ref{app:dS}, the constant of proportionality in eq. \eqref{eqn:dSapprox} is determined.

To compute the QES location, we need $\partial_x \delta S(\boldsymbol{\ell i}) = -2 \partial_{z_{\boldsymbol{\ell}}} \delta S(\boldsymbol{\ell i})$, where $z_{\boldsymbol{\ell}}$ is the endpoint of $\boldsymbol{\ell}$ in Poincare coordinates. By AdS$_2$ symmetry, moving the twist operator near the bifurcation surface is isometric to moving the near boundary twist operator instead. By scaling symmetry of AdS, we have 
\begin{align}
    \partial_{z_{\boldsymbol{\ell}}}S(\boldsymbol{\ell i}) =- \frac{z_{\boldsymbol{i}}}{z_{\boldsymbol{\ell}}} \partial_{z_{\boldsymbol{i}}}S(\boldsymbol{\ell i})
\end{align}
and so for $z_{\boldsymbol{\ell}}$ near $1$ (i.e. the bifurcation suface), we get that 
\begin{align}
S'(\boldsymbol{\ell i}) \sim -\left(z_{\boldsymbol{i}}\right)^{2\Delta}
\end{align}
where again the constant of proportionality is positive and only depends upon $\Delta$. Plugging this into the expression in eq. \eqref{eqn:dxS'}, and inputting the $\Delta$-dependent coefficient from Appendix \ref{app:dS}, we find
\begin{align}
    \delta x_Q = \frac{K \beta }{8 \pi \phi_r}\frac{\sqrt{\pi}}{4^{2\Delta}} \frac{2\Delta^2\Gamma(2\Delta)}{\Gamma(2\Delta + 3/2)} \left( \frac{z_{\boldsymbol{i}}}{2}\right)^{2\Delta}.
\end{align}

In order to compare this with the answer found from the boundary, eq. \eqref{eqn:mainqesresult}, one needs to have a prediction for $z_{\boldsymbol{i}}$. $z_{\boldsymbol{i}}$ should be thought of as the location of the ``QES'' for the $K$ flavors of boundary fermions. Since $K\ll N$, we expect that such a surface corresponds (via null light rays) to a boundary time band of order $\delta t \sim \frac{1}{\mJ}$. The constant of proportionality in this relation cannot be determined from the bulk, however. The reason is that for physical questions regarding observables too close to the boundary, we should not trust the bulk description of $N$ free fermions. The bulk description of SYK in terms of $N$ free fermions emerges only for time separations $\delta t/\beta \gg \frac{1}{\beta\mJ}$. A discussion of small intervals off near the boundary of AdS$_2$ dual to SYK also appeared in \cite{Qi:2021tz}.

Thus, we can really only compare the overall scaling with $z_{\boldsymbol{i}}$ and must ignore the overall coefficient. Indeed, ignoring the detailed dependence on $\Delta = 1/q$ of the overall $\mJ$-independent factor, we see agreement with eq. \eqref{eqn:mainqesresult}.

There is another way of testing this bulk picture which is insensitive to the coefficient in eq. \eqref{eqn:dSapprox}; we can compare the ratio of $\delta x_Q$ to $\delta S(\boldsymbol{\ell i})$. Via eqs. \eqref{eqn:dxS'} and \eqref{eqn:dSapprox}, we find that the bulk model predicts the precise relationship
\begin{align}\label{eqn:bulkdxdS}
    \delta x_Q = -\frac{\beta \Delta}{2\pi \phi_r} \delta S(\boldsymbol{\ell i}).
\end{align}
Assuming the bulk QES formula for the boundary entropy $S(LK)$, then we have that the leading order change to the entropy of $LK$ due to including $K$ is just
\begin{align}
    S(LK)-S(L) - S(K) = \delta S(LK) = \delta S(\boldsymbol{\ell i}).
\end{align}
In Appendix \ref{app:SYKEE}, we explicitly compute $\delta S(LK)$ in SYK and find
\begin{align}
    \delta S(LK) = -K b \frac{\pi^{2\Delta + 1/2}\Gamma(\Delta + 1)}{2\Gamma(\Delta + 3/2)},
\end{align}
with $b$ defined in eq. \eqref{eqn:bdef}. Comparing with the SYK prediction for $\delta x_Q$ in eq. \eqref{eqn:mainqesresult}, we find that $\delta x_Q$ and $\delta S(LK)$, as computed via SYK, obey eq. \eqref{eqn:bulkdxdS}.

\section{Discussion}\label{sec:discussion}
We now end with some brief comments on loose ends and future directions.
\subsection{Saturating the price bound}
In Sec. \ref{sec:pricesize}, we argued for a lower bound on the price, of the form 
\begin{align}
    p_{\beta}(O) \geq N-K_0
\end{align}
where
\begin{align}\label{eqn:Kstar}
    K_0 = \frac{N \beta \braket{H_R - H_L}_{O\ket{\beta}}}{2\pi S(T_R)}
\end{align}
and $S(T_R)$ is the operator size of $O_R(T_R) \rho_R^{1/2}$. This bound assumes that the operator $O_R(T_R) \rho_R^{1/2}$ displays detailed size-winding, as defined in Sec. \ref{sec:pricesize}. In the context of the SYK discussion, $O(T_R)\ket{\beta} = \psi(T_R+i\delta)\ket{\beta}$. The boost energy of this state at large $\beta \mJ$ is equal to
\begin{align}
    \frac{\beta}{2\pi} \braket{H_R-H_L}_{\psi \ket{\beta}} = 2\Delta  \frac{b \pi^{2\Delta} \cos \delta}{\sin^{1+2\Delta} \delta} 
\end{align}
with $\Delta = 1/q$ and $b$ defined in eq. \eqref{eqn:bdef}. We are also assuming $ \delta \gg \frac{1}{\beta \mJ}$. Plugging this into the eq. \eqref{eqn:Kstar} for $K_0$, and using the formulae in \cite{Qi:2018wg} to compute the size, $S(T_R)$, we get that
\begin{align}
K_0 \sim  \frac{N\alpha_s}{\beta \mJ} \frac{ \cos \delta}{b e^{-2\pi T_R}},
\end{align}
where we remind the reader that $b \sim \frac{1}{(\beta \mJ)^{2/q}}$.

As discussed in Sec. \ref{sec:pricecalc}, the price can be computed by solving the equation
\begin{align}
    \delta x^+_Q(K_*) = e^{2\pi T_R}
\end{align}
for $K_*$ and then the price is just 
\begin{align}
    p_{\beta}(\psi) = N- K_*.
\end{align}
Using the equation for $\delta x^+_Q$ derived in eq. \eqref{eqn:mainqesresult}, we find that $K_*$ obeys the equation
\begin{align}
    K_* \sim K_0
\end{align}
where we have ignored order $1$, $\Delta$-dependent factors. Thus, the price (or really $K_*$) parametrically saturates the bounds eqs. \eqref{eqn:Kbound} and \eqref{eqn:priceboundsize}.

This parametric saturation happens because the modular Hamiltonian $\log \Delta_{LK}$ is closely related to the size operator. In fact, our formula for the modular flowed correlators at large $\beta \mJ$ in sec. \ref{sec:analyticcontinuation} are consistent with the expression
\begin{align}\label{eqn:modhamLK}
    \log \rho_{LK} = \log \rho_L + \sum_{i=1}^K  \int_{-\infty}^{\infty} ds \frac{1}{4\cosh^2 \pi s} \psi^i_L(s) \psi_K^i + ...
\end{align}
where the ellipsis denote terms in the modular Hamiltonian which contribute to the modular flowed correlator at higher orders in $K/N$ and $1/(\beta \mJ)^{2/q}$ or do not contribute to the exponentially growing part of the correlator. Since $1/\cosh^2(\pi s)$ is fairly well peaked around $s=0$, we see that the correction to the modular energy from including $K$ is roughly proportional to the $K$-size operator, discussed in \cite{Schuster:2021uc, Brown:2019aa}.

\subsection{Emergent Type III von-Neumann algebra}

Our results for the modular flowed correlator at large $\beta \mJ$ illustrate that these sub-algebras associated to subsets of fermions in SYK obey a half-sided modular inclusion algebra (HSMI). A particularly interesting consequence of this HSMI is that it gives rise to null translations in the bulk, even though it itself is non-geometric (in the sense that it doesn't correspond to spatial inclusions). Explicitly, the form of the modular flowed correlator for the case of maximal modular chaos in eq. \eqref{eqn:maxmodchaos} implies the relationship
\begin{align}\label{eqn:HSMI}
    \Delta_{LK}^{is} \Delta_L^{-is} = \exp \left(i(e^{-2\pi s}-1) \delta x^+_Q P_+ \right)
\end{align}
with $\delta x_Q^+$ determined by eq. \eqref{eqn:mainqesresult}. By $P_+$ here, we mean a null shift generator, which moves operators along the horizon. This $P_+$ is one of the generators of the $SL(2, \mathbb{R})$ discussed by \cite{LMZ}. We have dropped a contribution from $P_-$ in eq. \eqref{eqn:HSMI} because we are only considering the action of these operators on highly boosted excitations. 

As discussed by \cite{Leutheusser:2021aa, Leutheusser:2021ab}, such an HSMI algebra can only be obeyed when the algebras $L$ and $LK$ are Type III$_1$, the type of algebra usually associated to a quantum field theory. SYK is not a quantum field theory and in fact has a finite Hilbert space of dimension $2^{N/2}$ at finite $N$. This means that the defining HSMI formula can only be true perturbatively in $1/N$, as predicted by \cite{Leutheusser:2021aa}. It would be interesting to analyze the finite $N$ effects which destroy the HSMI property.

Since SYK at large $N$ is effectively a generalized free field theory (GFF), one actually could have guessed the form of the modular Hamiltonian in eq. \eqref{eqn:modhamLK} just from GFF reasoning. As discussed by \cite{Witten:2021aa}, one expects a Gaussian term, since Wick's theorem applies to operators in $LK$, plus an operator which generates an outer automorphism (at large $N$) of $L$. Such an operator is the time evolution operator, dual in gravity to the area operator of the bifurcation surface. This is basically what we found in eq. \eqref{eqn:modhamLK}.

In this paper we considered subsets of fermions, but because SYK is a GFF at large $N$, one could have tried to compute the modular Hamiltonian for time bands, analogous to what was done in \cite{Leutheusser:2021aa,Leutheusser:2021ab}. Such a calculation would imply the existence of a QES for time bands, which one could explicitly try to find. We leave this for future work.

\subsection{Finite $\beta \mJ$ at large $q$}
As we explain in Sec. \ref{sec:analyticcontinuation}, the expression for the four point function is known at large $q$ and all $\beta \mJ$. It is interesting to compare the result for the modular energy at sub-maximal chaos, as given in eq. \eqref{eqn:modenergylargeq}, with that discovered in \cite{Chandrasekaran:2021wx}. In that work, the modular energy was computed for $v<1$ and was found to go from positive to negative at a time $T_R^*$ which depended on the smearing scale $\delta$. The authors in \cite{Chandrasekaran:2021wx} hypothesized that this effect corresponds to a stringy spreading of the probe as it falls across the quantum extremal surface. 

By contrast, we see that in SYK at large $q$ there is no $\delta$ dependence on the turnover time. If the hypothesis that $\delta$-dependence in $\delta x_Q$ is related to string spreading is correct, then we have found that the bulk dual for SYK at large $q$ is in some sense ``less'' stringy at sub-maximal chaos than it otherwise would be in higher-dimensions. We suspect that this lack of $\delta$-dependence is related to the change in the momentum wave-functions for SYK at large $q$, as discussed in Appendix D.2 of \cite{Nezami:2021td}. 

\subsection{Higher orders in $K/N$}
It would be interesting to compute the modular flowed correlator to higher orders in $\lambda = K/N$. In particular, it should be possible to work in a limit where $\lambda e^{-T_R}$ is small but where $e^{-T_R}$ is large. This will isolate a subset of the higher orders in $\lambda$ contribution. The naive guess is that the answer re-sums into the expression in eq. \eqref{eqn:allorderslambda}. 

This would require isolating contributions to the $G$-$\Sigma$ equations that are growing like $(\lambda e^{-T_R})^n$. Such contributions could perhaps be re-summed using the recent work of \cite{Gu:2021un}, but we leave this for future exploration.

\subsection{Quantum extremal surfaces in matrix quantum mechanics models?}
In this work, we found evidence for a qunatum extremal surface for a boundary algebra that is not associated to a spatial sub-region. This opens up the possibility that such a phenomenon occurs in other situations, such as in the (un-gauged) BFSS model of \cite{Maldacena:2018wa} where we could associate an algebra to a sub-set of matrix elements in $X$. Quantum error correction in $SU(N)$ matrix quantum mechanics was also discussed in the nice paper \cite{Milekhin:2021aa}, although the discussion was mostly kinematical. It would be interesting to understand the input from the dynamics of OTOCs in that context. Extremal surfaces in the bulk dual of BFSS were analyzed by \cite{Anous:2020aa}. Perhaps a modified analysis could find quantum extremal surfaces in the bulk as well.

\section*{Acknowledgements}
We thank Ahmed Almheiri, Lampros Lamprou, Nima Lashkari, Henry Lin, Juan Maldacena, Alex Streicher, Edward Witten, and Ying Zhao for helpful discussions. We particularly thank Henry Lin and Alex Streicher for helpful discussions regarding operator size and size winding in SYK. A.L. acknowledges support from NSF grant PHY-1911298 and Carl P. Feinberg. V.C. is supported by a grant from the Simons Foundation (816048, VC).

\appendix 
\section{Perturbing by a fermion in $K$}\label{app:Kfermion}
Throughout the main text, we considered exciting the TFD by a single fermion, which for convenience we took to be one of the $N-K$ fermions in $r$. This choice was purely for convenience, and in this appendix we show that we get the same prediction for the QES position $\delta x_Q$ if we excite the TFD with a fermion in $K$ instead.

If we excite the TFD with a fermion in $K$, then the relevant modular flowed correlator to compute the QES position is
\begin{align}
    \braket{\Delta_{LK}^{is}}_{\psi_{\delta}} = \frac{Z_{\delta}^{-1}}{K}\sum_{i=1}^{K} \braket{\psi_{\delta}^i\Delta_{LK}^{is}\psi_{\delta}^i}_{\beta}.
\end{align}
This can be found by solving for the two point function $G_K$ with the boundary conditions laid out in eq. \eqref{eqn:bcs}. As we did for the correlator in the main text, we need to solve for $G_K^{(k,0)}(\tau,\tau')$ at integer $n$ and $k \geq n-1$ and then continue $n \to 1$ and $k \to is$. 

To solve for $G_K^{(j,k)}(\tau, \tau')$, we return to the $G$-$\Sigma$ equations
\begin{align}
&\partial_{\tau} \mathbf{G_K}(\tau, \tau') - \left(\mathbf{\Sigma_r}\star \mathbf{G_K}\right)(\tau, \tau') = \mathbf{1} \delta (\tau - \tau'),\nonumber \\
&\Sigma^{(j,k)}_K = \Sigma_r^{(j,k)} =  J^2 \left(G_r^{(j,k)}\right)^{q-1} + J^2 \lambda (q-1) \left(G_K^{(j,k)} - G_r^{(j,k)}\right) \left(G_r^{(j,k)}\right)^{q-2}+  \mathcal{O}(\lambda^2),
\end{align}
where we remind the reader that $\mathbf{G_K}$ is a matrix in replica indices. It is useful to introduce a function $G_K$ (without super-scripts) that is a function of two times which both live on an $n\beta$-length circle. This function $G_K(\tau, \tau')$ was defined in eq. \eqref{eqn:GKnbeta}. 

In terms of this function, the Schwinger-Dyson equations for the order $\lambda$ piece of $G_K$ is
\begin{align}
    \left(\partial_{\tau}- \Sigma\right) \star G_{K,1} -  (q-1) J^2 G^{q-2} G_{r,1} \star G_{K,0} - J^2(q-1) G^{q-2} (G_K - G_r) \star G_{K,0}= \sum_{j} \delta(\tau - j) \disc G_{K,1}(j,\tau').
\end{align}
Substituting in the formula for $G_{K,0}$ we derived in the main text
\begin{align}
    G_{K,0}(\tau, \tau') = G(\tau, \tau') + \sum_{\ell m} G(\tau,\ell)c_{\ell m} G(m,\tau')
\end{align}
we get the expression
\begin{align}
    &\left(\partial_{\tau}- \Sigma\right) \star \left(G_{K,1} -G_{r,1}\right)\nonumber \\
    &= (q-1) J^2 G^{q-2} G_{r,1} \star (G \star c \star G) + J^2(q-1) G^{q-2} (G_{K,0} - G) \star (G \star c \star G)+ \sum_{j} \delta(\tau - j) \disc G_{K,1}(j,\tau').
\end{align}
We have introduced the short hand
\begin{align}
   \left( G \star c \star G\right) (\tau, \tau') \equiv \sum_{\ell m} G(\tau, \ell) c_{\ell m} G(m,\tau') .
\end{align}

As we did in the main text, we can matrix multiply $G$ in from the left on both sides of the equality and we find
\begin{align}
    G_{K,1} = G_{r,1} + \frac{1}{2}(q-1) J^2 G \star G^{q-2} (\mF \star \star\, c) \star (G \star c \star G) + \sum_j G(\tau, j) \disc G_{K,1}(j,\tau')
\end{align}
where we have also introduced the notation
\begin{align}
    (\mF \star \star\, c)(\tau, \tau') =\sum_{\ell m} \mF(\tau, \tau'; \ell, m) c_{\ell m}
\end{align}
and so is a two-index tensor (matrix).
Using the definition for the ladder kernel $K_c$ defined in eq. \eqref{eqn:Kc}, we can write this equation as
\begin{align}
    G_{K,1} = G_{r,1} + \frac{1}{2} K_c \star (\mF \star \star\, c) \star c \star G + \sum_j G(\tau, j) \disc G_{K,1}(j,\tau').
\end{align}
Using that $K_c \star \mF = \mF - \mF_0$ and unpacking this notation we find
\begin{align}
    G_{K,1}(\tau, \tau') = G_{r,1}(\tau, \tau') + G_{r,1}(\tau, \ell) c_{\ell m} G(m, \tau')+ G(\tau, j) \disc G_{K,1}(j,\tau').
\end{align}
where repeated indices are summed over. Now we need to impose boundary conditions. It is clear that to preserve the anti-symmetry of $G_{K,1}$ in exchanging $\tau$ and $\tau'$, we need
\begin{align}
    \disc G_{K,1}(j,\tau') = \sum_m c_{jm}G_{r,1}(m,\tau') + \beta_{jm} G(m,\tau')
\end{align}
where $\beta_{jm}$ is a new matrix which in principle is determined by the remaining boundary condition. The remaining boundary condition is that $G_{K,1}$ live on $n$ copies of a $\beta$-length circle. We can try to solve this boundary condition for $\beta_{\ell m}$ at large $\beta \mJ$. We can remember, however, that $\mF$ is enhanced relative to $G$ by factors of $\beta \mJ$ and so we can actually drop the term involving $\beta_{\ell m}$ to leading order.

Cyclicity then gives the constraint on $c_{\ell m}$ at large $\beta \mJ$
\begin{align}
    &G_{r,1}(j+1^-,\tau') + G_{r,1}(j^+,\tau') \nonumber \\
    &= -\sum_{\ell m} \left(G(j+1^-,\ell) + G(j^+,\ell)\right)c_{\ell m} G_{r,1}(m,\tau') - \sum_{\ell m} \left( G_{r,1}(j+1^-,\ell) + G_{r,1}(j^+,\ell)\right) c_{\ell m}G(m,\tau').
\end{align}
Now, for $\tau'$ away from an integer, the second term on the right hand side is suppressed by $1/(\beta \mJ)^{2/q}$ relative to the terms on the left where $\ell = j+1$ or $\ell = j$. Keeping only these terms, we find the constraint
\begin{align}
    &G_{r,1}(j+1^-,\tau') + G_{r,1}(j^+,\tau') \nonumber \\
    &= -\sum_{m} \frac{1}{2} \left(c_{j m} - c_{j+1 m}\right) G_{r,1}(m,\tau').
\end{align}
We find the equation
\begin{align}
    \frac{1}{2} \left(c_{\ell+1 m} - c_{\ell m}\right) = \delta_{j+1 m} + \delta_{j m},
\end{align}
which is the same equation for $c_{\ell m}$ that we found in eq. \eqref{eqn:clmlargebetaJ} for the large $\beta \mJ$ expansion coefficients. For $\tau$, $\tau'$ away from the integers, these extra terms in $G_{K,1}$ are actually suppressed at large $\beta J$. Thus, we find the equation
\begin{align}
    G_{K,1}(\tau, \tau') = G_{r,1}(\tau, \tau')\left(1+ \mathcal{O}(\frac{1}{(\beta \mJ)^{2/q}})\right),
\end{align}
and so we find that it does not matter whether we use a fermion in the $r$ subset or the $K$ subset to compute the price at large $\beta \mJ$. It would be interesting to understand the difference at large $q$ and finite $\beta \mJ$ but we leave that for future work.

\section{Free Majorana Fermion Propagators in $AdS_2$}\label{app:bulkfermion}
The discussion of Majorana fermions in AdS$_2$ we detail here is based upon the discussion in the Appendix of \cite{Beccaria:2019tm}. Majorana fermions in AdS$_2$ have the action
\begin{align}
S = \int \frac{dt dz}{z^2} \left(z \psi \overline{\partial} \psi + z \overline{\psi} \partial \overline{\psi} - im \overline{\psi} \psi \right).
\end{align}
Introducing re-scaled fermions $\psi = z^{1/2} \eta$ and $\bar{\psi} = z^{1/2} \bar{\eta}$, we get the action
\begin{align}
S = \int dt dz \left( \eta \bar{\partial} \eta + \bar{\eta} \partial \bar{\eta} - \frac{im}{z} \bar{\eta} \eta \right).
\end{align}
Using complex Poincare coordinates, $w = t +iz$, we find the equations of motion 
\begin{align}
\bar{\partial}\eta -\frac{m}{w - \bar{w}} \bar{\eta} = 0, \ \ \partial \bar{\eta} + \frac{m}{w-\bar{w}} \eta = 0
\end{align}
Differentiating again leads to the equations
\begin{align}
&\left(\partial \bar{\partial} \eta  + \frac{m}{(w-\bar{w})^2} \bar{\eta}  - \frac{m}{(w-\bar{w})} \partial \bar{\eta}\right) \nonumber \\
& = \left(\partial \bar{\partial} \eta  + \frac{1}{w-\bar{w}} \bar{\partial} \eta  + \frac{m^2}{(w-\bar{w})^2} \eta\right) = 0.
\end{align}

We can then solve for the propagator $\braket{ \eta(w, \bar{w}) \eta (w', \bar{w}')}$ and find \cite{Beccaria:2019tm}
\begin{align}
\braket{ \eta(w, \bar{w}) \eta (w', \bar{w}')} = \frac{\wb - \wb'}{(w- \wb)(w'-\wb')} F(u)
\end{align}
where $u = \frac{(z-z')^2 + (t-t')^2}{2zz'} = -2 \frac{(w-w')(\wb - \wb')}{(w-\wb)(w'-\wb')}$ and
\begin{align}
&F(u) \nonumber \\
&= c_1 \left(\frac{2}{2+u} \right)^{m+1} {}_2F_1\left(m+1, m+1; 2m+1; \frac{2}{2+u}\right) + c_2  \left(\frac{2}{2+u} \right)^{1-m} {}_2F_1\left(1-m, 1-m;1-2m;\frac{2}{2+u}\right).
\end{align}

For us, $m=-\frac{1}{2} + \frac{1}{q}$. We can find what the coefficients need to be by looking at the limit where one of the points goes to the boundary. This is the limit where $u \to \infty$. In this case, both terms decay, although one faster than the other. A natural normalization is to set $c_2 = 0$, $m = -\frac{1}{2} + \frac{1}{q}$. We then get 
\begin{align}
&\braket{\psi(w,\wb) \psi(w',\wb')}
& = c_1 \frac{\wb - \wb'}{\sqrt{-(w- \wb)(w'-\wb')}}   \left(\frac{2}{2+u} \right)^{m+1} {}_2F_1\left(m+1, m+1; 2m+1; \frac{2}{2+u}\right) \nonumber \\
&\braket{\psi(w,\wb) \bar{\psi}(w',\wb')},
& = c_1 \frac{-(w - \wb')}{\sqrt{-(w- \wb)(w'-\wb')}}   \left(\frac{2}{2+u} \right)^{m+1} {}_2F_1\left(m+1, m+1; 2m+1; \frac{2}{2+u}\right).
\end{align}
Moving both points near the boundary gives
\begin{align}
\braket{\psi(w,\wb) \psi(w',\wb')} = 4^{m+1/2} (z z')^{m+1/2} \frac{c_1}{(t-t')^{2m+1}}.
\end{align}

\subsection{Coordinates for AdS$_2$}\label{app:coords}
We have the two point function in Poincare coordinates but we would like the two point function in hyperbolic coordinates instead. In these coordinates, the metric is instead
\begin{align}
ds^2 = d\rho^2 + \sinh^2(\rho) d\tau^2.
\end{align}

We can describe these two coordinate systems using embedding space. For Euclidean $AdS_2$, we have
\begin{align}
-Y_{-1}^2 + Y_0^2 + Y_1^2 = -1.
\end{align}
To relate Poincare to hyperbolic coordinates, we use
\begin{align}
Y_{-1} = \frac{z}{2}(1 + \frac{1}{z^2}(1+t^2)) = \cosh(\rho),\ Y_0 = \frac{t}{z} = \sinh(\rho) \sin \tau,\ Y_1 = \frac{z}{2}(1- \frac{1}{z^2}(1-t^2)) = \sinh(\rho) \cos \tau
\end{align}
This relation gives
\begin{align}
\tan(\tau) =  \frac{2t}{z^2+t^2 -1}.
\end{align}
From this we see that $\tau \in (-\pi/2, \pi/2)$. 
Furthermore, 
\begin{align}
\cosh(\rho) = \frac{z}{2}(1+ \frac{1}{z^2}(1+t^2)).
\end{align}

The isometries of AdS just act as the isometries of the ambient $\mathbb{R}^{1,2}$, and so the invariant AdS distance between two points is just the distance between those two points in the embedding space. In other words, the invariant AdS distance $u$ is given by 
\begin{align}
u = \frac{(t-t')^2 + (z-z')^2}{2zz'} = \cosh(\rho) \cosh(\rho') -\cos(\tau - \tau') \sinh(\rho) \sinh(\rho')-1.
\end{align}
 
\section{Computing $\delta S(\boldsymbol{\ell i})$}\label{app:dS}
To compute $S(\bl \bi)$, we just need to compute the entanglement entropy for some arbitrary region $\bl \cup \bi$ where $\bi$ is far out near the boundary. This involves inserting two twist operators, one near the bifurcation surface and one out near the boundary. This amounts to computing the partition function of free fermions in the presence of branch points. Similar computations were done in \cite{Agon:2015ftl} and we follow their lead. 

The idea is to expand the twist operators into a complete set of operators in the bulk theory. When we expand both twist operators, the leading contribution is from bi-replica operators. See \cite{Agon:2015ftl} for more details. Dropping everything else since it will be sub-leading in the large $\beta \mJ$ limit, the partition functions in the presence of the conical defects becomes 
\begin{align}
\frac{Z_n(C_{\bl} \cup C_{\bi})Z^n}{Z_n(C_{\bl}) Z_n(C_{\bi})} = 1 + \frac{1}{2} \sum_{j \neq j' } C_{j,j',\alpha \beta}^{\bi} \braket{\psi_j^{\alpha}(r_0)\psi_{j'}^{\beta}(r_0)}_{C_{\bl}}+...\ .
\end{align}
The Greek indices here denote the different chiralities of the field i.e. $\psi$ and $\bar{\psi}$. Here by $r_0$ we mean the point in AdS$_2$ given by $r_0 = (\tau = 0, \rho = \rho_0)$ with $\rho_0$ large. Note that this point is arbitrary in this calculation and so dependence on it should cancel out in the end. From here, it will be convenient to change field variables to $\psi^{\pm} = \psi \pm \bar{\psi}$. This is because $\braket{\psi_j^{-} \psi_{j'}^{+}} = \braket{\psi_j^{+} \psi_j^{+}} = 0$ in the large $\rho$ limit. Thus, we get
\begin{align}\label{eqn:operatorexpansion}
&\frac{Z_n(C_{\bl} \cup C_{\bi})Z^n}{Z_n(C_{\bl}) Z_n(C_{\bi})} =1 + \frac{1}{2} \sum_{j \neq j' } C_{j,j',- -}^{\bi} \braket{\psi_j^{-}(\tau = 0, \rho_0) \psi_{j'}^{-}(\tau = 0, \rho_0)}_{\mathcal{C}_{\bl}} +...\ .
\end{align}

Now we need to compute the correlation functions and the OPE coefficients. The OPE coefficients can themselves be computed in terms of the two point function in the presence of a twist operator. In particular, to get $C^{\bi}_{j,j',--}$ we consider probe operators at the AdS$_2$ point $r = (\tau = \pi/2,\rho)$ and take $\rho$ large. The choice of $\tau = \pi/2$ is for convenience. Then we have
\begin{align}
\lim_{\rho \to \infty} \braket{\psi^{(j)-} (\pi/2, \rho) \psi^{(j')-}(\pi/2, \rho) }_{\mathcal{C}_{\bi}} = \lim_{\rho \to \infty} C^{\bi}_{j,j',--}G^{--}(r,r_0)G^{--}(r,r_0)
\end{align}
where in this equality, we have used that $G^{\pm \mp}(r,r_0) = G^{++}(r,r_0) = 0$. At the point $\tau = \pi/2$, the relation between Poincare and hyperbolic coordinates is $z = 1/\cosh(\rho)$ and $t = \tanh(\rho)$, as shown in Appendix \ref{app:coords}. Using the propagator formulae in appendix \ref{app:bulkfermion}, we have
\begin{align}
G^{--}(r,r_0) = 4 G^{\psi \psi}(r,r_0) = 4c_1 \left( 4zz_0\right)^{m+1/2}
\end{align}
with $z = 2e^{-\rho}$ and $z_0 = 2e^{-\rho_0}$. In the end, we find
\begin{align}
C_{j,j',--}^{\bi} = 
&\frac{\braket{\psi^{-,(j)}(\pi/2,\rho) \psi^{-,(j')} (\pi/2,\rho)}_{\mathcal{C}_{\bi}}}{16 c_1^2(4 zz_0)^{2m+1}}.
\end{align}

Now we need to compute the two-point function in the presence of twist operators. We start with the computation of $\braket{\psi_j^{-}(\tau = 0, \rho_0) \psi_{j'}^{-}(\tau = 0, \rho_0)}_{\mathcal{C}_{\bl}}$. Following the lead of \cite{Agon:2015ftl}, we compute this two point function for general values of $\tau, \tau'$. Namely, we would like to compute $\braket{\psi_j^{-}(\tau, \rho_0) \psi_{j'}^{-}(\tau', \rho_0)}_{\mathcal{C}_{\bl}}$, which amounts to computing the two-point function in a background with a conical excess. To do this, we instead map the replica index $n \to n= 1/m$ and work on a background with a conical deficit, with opening angle $2\pi /m$. We can then compute the two point function on this conical deficit background via the method of images for integer $m$. Having done this, we then continue this quantity below $m =1$ to $m = 1/n$ for integer $n$, plug the resulting answer back into eq. \eqref{eqn:operatorexpansion} and continue that back down to $n =1$.

To find the two point function in the presence of a conical deficit, we use
\begin{align}
&\braket{\psi^-(\tau,\rho_0) \psi^-(\tau',\rho_0)}_{AdS_2}  \nonumber \\
& = 4c_1 4^{m+1/2} e^{-(2m+1) \rho_0} \left( \frac{1}{\sin((\tau - \tau')/2)}\right)^{2m+1}
\end{align}
where we have used the formulae in Appendix \ref{app:bulkfermion} and Appendix \ref{app:coords} to write the two point function in hyperbolic coordinates. Using the method of images, we then find that 
\begin{align}\label{eqn:Clorbifold}
    \braket{\psi^-(\tau, \rho_0) \psi^{-}(\tau', \rho_0)}_{\mathcal{C}^{1/m}_{\ell}} =  c_1 4^{2m+2} e^{-(2m+1) \rho_0} g_m(\tau - \tau'),
\end{align}
where
\begin{align}
    g_m(\tau) = \sum_{k=0}^{m-1} \left(\frac{1}{4\sin^2(\pi k/m + \tau)}\right)^{m+1/2}.
\end{align}
To get $\braket{\psi_j^-(\tau = \pi/2,\rho) \psi_{j'}^-(\tau = \pi/2,\rho)}_{\mathcal{C}_{\bi}}$, we can use the AdS$_2$ isometries to move the twist operator associated to $\boldsymbol{i}$, which sits at $r_I = (\tau = 0, \rho_{\bi})$, to the origin in AdS$_2$. Note that the $\tau$ coordinates of the operators will be translated to near $\tau = \pi$. We can also use the AdS isometries to see that $\rho \to \rho + \rho_{\bi}$ for large $\rho$. Thus, we have that
\begin{align}
    \braket{\psi_j^-(\tau = \pi/2,\rho) \psi_{j'}^-(\tau' = \pi/2,\rho)}_{\mathcal{C}_{\bi}} = \braket{\psi_j^-(\tau \approx \pi,\rho+\rho_{\bi}) \psi_{j'}^-(\tau' \approx \pi, \rho+\rho_{\bi})}_{\mathcal{C}_{\bl}}.
\end{align}
Plugging in the answer from eq. \eqref{eqn:Clorbifold} for this two point function, we get
\begin{align}
&\frac{Z_n(C_A \cup C_B)Z^n}{Z_n(C_A) Z_n(C_B)} \nonumber \\
&=1 + \frac{1}{2}\sum_{j \neq j' } e^{-(2m+1)\rho_{\bi}} \left(  g_{1/n}(2\pi (j-j')) \right)^2
\end{align}
where $g_{1/n}(j)$ is the analytic continuation of $g_m(\tau)$ to $m=1/n$. After continuing in $m$, we can also set $\tau \to \tau+ 2\pi j$ for $\psi_j(\tau)$ and $\tau \to \tau' + 2\pi j'$ for $\psi_j(\tau')$.

To finish the computation, we see that we just have to analytically continue the sum
\begin{align}
\sum_{j \neq j' } \left(  g_{1/n}(2\pi (j-j')) \right)^2
\end{align}
to $n=1$. This is the same sum that was continued in Agon \& Faulkner up to the replacement $2m+1 \to 2\Delta$ \cite{Agon:2015ftl}. Taking their expression, we get
\begin{align}
&\lim_{n\to 1} \frac{1}{n-1} \sum_{j \neq j' } \left(  g_{1/n}(2\pi (j-j')) \right)^2 = \frac{2\sqrt{\pi}}{4^{2m+2}} \frac{\Gamma(2m+2)}{\Gamma(2m+5/2)},
\end{align}
and the entropy is $S(\ell I ) = S(\boldsymbol{\ell}) + S(\boldsymbol{i}) +\delta S(\boldsymbol{\ell i})$ where
\begin{align}
 \delta S(\boldsymbol{\ell i}) = -\frac{\sqrt{\pi}}{4^{2m+2}} \frac{\Gamma(2m+2)}{\Gamma(2m+5/2)} \left( \frac{z_{\boldsymbol{i}}}{2}\right)^{2m+1},
\end{align}
where we rewrote the position of the endpoint of $\boldsymbol{i}$, given by $\rho_{\bi}$, in Poincare coordinates point $z_{\boldsymbol{i}} \sim \frac{1}{\beta \mJ}$, with a constant of proportionality that needs to be fixed.

\section{SYK twist operators}\label{app:SYKtwistop}
In the main text, we computed the modular flowed correlator eq. \eqref{eqn:modflowcorr} using the $G$-$\Sigma$ equations. There is an equivalent method, however, using twist operators, which we detail in this appendix. In particular, the correlator
\begin{align}\label{eqn:modflowcorrtwist}
    \braket{\psi \rho_{LK}^p \otimes \rho_{r}^{n-1-p} \psi}_{\beta} = \braket{\psi^{(p)} \sigma_{L}^n \otimes \sigma_K^{-n} \psi^{(0)}}_{\beta^{\otimes n}}
\end{align}
where $\sigma_L^n$ is a twist operator acting on the Hilbert space $\mathcal{H}_L^{\otimes n}$ and $\sigma_K^{-n}$ is an anti-twist operator acting on the Hilbert space $\mathcal{H}_K^{\otimes n}$. These operators are defined in terms of the replicated theory as obeying the condition
\begin{align}
    &\sigma_L^n \psi_L^{(j)} = \psi_L^{(j+1)}\sigma_L^n \nonumber \\
    &\sigma_K^{-n} \psi_K^{(j)} = \psi_K^{(j+1)}\sigma_K^{-n},
\end{align}
so that it connects the $j$'th replica to the $j+1$'st replica.

These twist operators can then be expanded into a complete set of operators acting on $\mathcal{H}_{SYK}^{\otimes n}$. In particular, we have that
\begin{align}\label{eqn:twistopexpansion}
    \sigma_K^{-n}\ket{\beta}^{\otimes n} =\left(1 + \frac{1}{2} \sum_{j,\alpha;\  (j_1,\alpha_1) \neq (j_2, \alpha_2)} C^{\alpha_1, \alpha_2}_{j_1,j_2} \psi_{\alpha_1}^{(j_1)} \psi_{\alpha_2}^{(j_2)} +...\right)\ket{\beta}^{\otimes n},
\end{align}
where the $\alpha$'s are flavor indices in $K$ which run from $\alpha =1,...,K$ and $j_1, j_2$ are replica indices. The factor of $1/2$ is a symmetry factor so that we do not overcount operators in the expansion. The dots denote higher fermion number operators. If we plug this ansatz into eq. \eqref{eqn:modflowcorrtwist}, we find that at leading order in $1/N$, the answer is of the form
\begin{align}
    G_K(\tau, \tau') =G_{n\beta}(\tau - \tau') + \sum_{\ell, m} G_{n\beta}(\tau-\ell) c_{\ell m} G_{n\beta}(m-\tau') + \mathcal{O}(1/N),
\end{align}
where the $c_{\ell m}$ account for contributions for higher fermion operators in eq. \eqref{eqn:twistopexpansion}. This is why we guessed the ansatz for $G_K$ around eq. \eqref{eqn:GKansatz}.

We can furthermore reproduce our results for the large $\beta \mJ$ answer of the modular flowed correlator at integer $n$. To leading order in $1/N$, we can ignore contractions across fermion flavor and so we can find the coefficients in eq. \eqref{eqn:twistopexpansion} by computing, for example,
\begin{align}
    C_{j_1,j_2}^{\alpha_1, \alpha_2} = 4\braket{\sigma_K^{-n}\psi_{\alpha_2}^{(j_2)} \psi_{\alpha_1}^{(j_1)} }_{\beta^{\otimes n}} + \mathcal{O}(1/N) = \text{Tr} \left(\mathcal{T}\rho_K^n \left(\rho_K^{-j_1} \psi_{\alpha_1}\rho_K^{j_1} \right) \left(\rho_K^{-j_2} \psi_{\alpha_2} \rho_K^{j_2} \right)\right),
\end{align}
where the $\mathcal{T}$ denotes Euclidean time ordering with respect to $\rho_K$ evolution. The correlator on the right hand side is just a replica two point function on a manifold with the opposite boundary conditions to what we have considered in the main text. This two point function was computed in \cite{Zhang:2020us} to leading order in $K/N$. They found that, to leading order in $K/N$, $\rho_K$ is just maximally mixed so that its time evolution is trivial. In other words, we have
\begin{align}\label{eqn:twofermionopecoeff}
    C^{\alpha_1,\alpha_2}_{j_1,j_2} = \delta^{\alpha_1, \alpha_2} 2\,\sgn(j_1-j_2).
\end{align}

Plugging this into eq. \eqref{eqn:modflowcorrtwist} and dropping the higher fermion number contribution, we see agreement with the answer in eq. \eqref{eqn:GrlargebetaJ}. The reason that we can drop the higher fermion number operators is that they will always involve contractions between replicas. At large $\beta \mJ$ and fixed $1/q$, these cross-replica contractions will always come with factors of $\left( \frac{1}{\beta \mJ}\right)^{1/q}$.

This point also explains why the large $q$ answer for $c_{\ell m}$ in eq. \eqref{eqn:Gr1largeq} is different from that at large $\beta \mJ$ but fixed $q$; the large $q$ answer contains contributions from higher fermion number operators in eq. \eqref{eqn:twistopexpansion}. Furthermore, this also explains why our large $q$ answer in eq. \eqref{eqn:Gr1largeq} could not be analytically continued in $n,\,p$ easily; These higher fermion number operators also involve higher-replica number operators as well, whose contribution is hard to analytically continue. The contribution of these higher-replica operators suggests that the modular flow at large $q$ does not correspond to a nice local form, as it does at large $\beta \mJ$ and fixed $q$.

\section{Entanglement entropy of $LK$ at large $\beta J$}\label{app:SYKEE}

Using this twist operator perspective, it is not hard to also write down the leading order in $K/N$ change to the entanglement entropy associated to $LK$. The entanglement entropy of $LK$ can be computed by computing $\braket{\sigma_L^n \sigma_K^{-n}}_{\beta^{\otimes n}}$. As discussed in the previous subsection, at large $\beta J$ all the higher fermion number contributions are suppressed. Thus, we can restrict just to the two fermion contribution, where the two-fermion coefficient was found in eq. \eqref{eqn:twofermionopecoeff},
\begin{align}
    C_{\ell m} = 2\sgn(\ell - m).
\end{align}
Putting in the expansion in eq. \eqref{eqn:twistopexpansion}, we get
\begin{align}\label{eqn:mutualinfo}
    \frac{\braket{\sigma_L^n \sigma_K^{-n}}_{\beta^{\otimes}}\braket{\beta|\beta}^n}{\braket{\sigma_L^n}_{\beta^{\otimes n}}\braket{\sigma_K^{-n}}_{\beta^{\otimes n}}} =\frac{\braket{\beta | \beta}^{n}}{\braket{\sigma_K^{-n}}_{\beta^{\otimes n}}}\left( 1 + 2K \sum_{\ell > m} G_{\beta n}(\ell-m) + \mathcal{O}(1/(\beta \mJ)^{4/q})\right).
\end{align}
Taking the log of the left hand side of eq. \eqref{eqn:mutualinfo} and continuing $n \to 1$, we get the mutual information between $L$ and $K$. Thus, we need to continue the sum on the right hand side of eq. \eqref{eqn:mutualinfo}. This can be done as in the main text. We have
\begin{align}
    2\sum_{\ell>m} G_{\beta n}(\ell - m) = n \sum_{k=1}^n G_{\beta n}(k).
\end{align}
Following the methods in Sec. \ref{sec:analyticcontinuation}, we get 
\begin{align}
    S_{\beta}(LK) = S_{\beta}(L)+S_{\beta}(K)+\delta S_{\beta}(LK) + \mathcal{O}(1/(\beta \mJ)^{4/q}),
\end{align}
with 
\begin{align}
   \delta S_{\beta}(LK) \equiv -\frac{\pi}{2} K \int_{-\infty}^{\infty}ds \frac{G_{\beta}(-is+1/2)}{\cosh^2 \pi s}.
\end{align}
Using the form of the thermal two point function in eq. \eqref{eqn:Gn=1} and perfoming the $s$ integral we get 
\begin{align}
   \delta S_{\beta}(LK) =- K b \frac{\pi^{2\Delta + 1/2}\Gamma(\Delta + 1)}{2\Gamma(\Delta + 3/2)},
\end{align}
where we remind the reader that $b$ obeys the equation
\begin{align}
    (\beta J)^2 b^q \pi = \left( \frac{1}{2}-\Delta \right) \tan \pi \Delta,\ \ \Delta = \frac{1}{q},
\end{align}
so that $b$ scales as $1/(\beta J)^{2/q}$.

\bibliographystyle{JHEP}
\bibliography{all}
\end{document}